\documentclass[prd,
showpacs,preprintnumbers,amsmath,amssymb,
superscriptaddress,
a4paper,nofootinbib,longbibliography]{revtex4-2}

\usepackage{changes}

\usepackage{color}
\usepackage{natbib}
\usepackage{graphicx}% Include figure files
\usepackage{epsf,epsfig,float,latexsym,amsthm,fancyhdr,rotating}
\usepackage{graphics,psfrag,longtable}
\usepackage{slashed}

\usepackage{esint}
\usepackage{multirow}
\usepackage{array}
\usepackage{rotating}
\usepackage[colorlinks,citecolor=blue,linktoc=all,linkcolor=blue,urlcolor=blue]{hyperref}

\usepackage{mathrsfs} 

\usepackage{nicefrac}

\usepackage{upgreek}
\usepackage{multirow}

\usepackage{hhline}

\def\beq{\begin{equation}}
\def\eeq{\end{equation}}
\def\bea{\begin{eqnarray}}
\def\eea{\end{eqnarray}}
\def\beqa{\begin{equation}\begin{array}{l}}
\def\eeqa{\end{array}\end{equation}}
\def\eqlab#1{\label{eq:#1}}
\def\figlab#1{\label{fig:#1}}

\def\seclab#1{\label{sec:#1}}
\def\eref#1{(\ref{eq:#1})}
\def\Eqref#1{Eq.~(\ref{eq:#1})}

\def\Figref#1{Fig.~\ref{fig:#1}}

\def\secref#1{Sec.~\ref{sec:#1}}
\def\half{\mbox{$\frac{1}{2}$}}

\def\quarter{\mbox{$\frac{1}{4}$}}

\def\barr{\left(\begin{array}{c}}
\def\earr{\end{array}\right)}
\def\bmat{\left(\begin{array}{cc}}
\def\emat{\end{array}\right)}
\def\al{\alpha}

\def\ga{\gamma} 
 \def\De{\Delta}

\def\si{\sigma}

\def\nn{\nonumber}
\def\dd{\mathrm{d}}

\def\bq{\boldsymbol{q}}

\DeclareMathOperator\arccoth{arccoth}
\DeclareMathOperator\im{Im}
\def\3d{3-D}

\def\ol#1{\overline{#1}}

\def\bq{\mathbf{q}}

\def\2PE{2$\upgamma$}

\begin{document}
\title {Chiral perturbation theory of the hyperfine splitting in (muonic) hydrogen}
\author{Franziska~Hagelstein} 
\email{hagelste@uni-mainz.de}
\affiliation{Institute  of  Nuclear  Physics,
Johannes Gutenberg Universit\"at Mainz, 55099 Mainz, Germany
}
\affiliation{PRISMA+ Cluster  of  Excellence, Johannes  Gutenberg  Universit\"at  Mainz, 55099 Mainz, Germany}
\affiliation{Laboratory for Particle Physics, Paul Scherrer Institute, 5232 Villigen PSI, Switzerland 
}
\author{Vadim Lensky}
\author{Vladimir Pascalutsa}
\affiliation{Institute  of  Nuclear  Physics,
Johannes Gutenberg Universit\"at Mainz, 55099 Mainz, Germany
}

\begin{abstract}

The ongoing experimental efforts to measure the
hyperfine transition in muonic hydrogen prompt 
an accurate evaluation of the proton-structure effects. At the leading order in $\alpha$, which is $O(\alpha^5)$ in the hyperfine splitting (hfs),  these effects are usually evaluated in a data-driven fashion, using the empirical information on the proton electromagnetic form factors and spin structure functions. Here we perform a first calculation based on the baryon chiral perturbation theory (B$\chi$PT). At leading orders it provides  a prediction for the proton polarizability effects in hydrogen (H) and
muonic hydrogen ($\mu$H).
We find large cancellations among the various  contributions leading to, within the uncertainties, a zero polarizability effect at leading order in the B$\chi$PT expansion. 
This result is in significant disagreement 
with the current data-driven evaluations.  The small polarizability effect implies a smaller Zemach radius $R_\mathrm{Z}$, if one uses the well-known  experimental $1S$ hfs in H or the $2S$ hfs in $\mu$H. We, respectively, obtain $R_\mathrm{Z}(\mathrm{H}) = 1.010(9)$ fm,  $R_\mathrm{Z}(\mu\mathrm{H}) = 1.040(33)$ fm. The total proton-structure effect to the hfs at $O(\alpha^5)$ is then consistent with previous evaluations;
the discrepancy in the polarizability is compensated by the smaller Zemach radius.
Our recommended value for the $1S$ hfs in $\mu\text{H}$  is $182.640(18)\,\mathrm{meV}.$
\end{abstract}
%\pacs{}
\date{\today}
\maketitle
\begin{small}
\tableofcontents
\end{small}
\newpage

\section{Introduction}
Muonic-atom spectroscopy has been successful at determining the charge radii of proton, deuteron, helion and alpha-particle with unprecedented precision through Lamb shift measurements \cite{Pohl:2010zza,CREMA:2016idx,Krauth:2021foz}. It also holds the potential to impact tests of ab-initio nuclear theories and bound-state QED \cite{Antognini:2022jec}.
The proton
Zemach radius $R_\mathrm{Z}$ has been extracted from a measurement of the $2S$ hyperfine splitting (hfs) in muonic hydrogen ($\mu$H) \cite{Antognini:1900ns} with a $3.4\,\%$ uncertainty:
\beq
R_\mathrm{Z}(\mu\mathrm{H})=1.082(37)\, \mathrm{fm},\eqlab{RZmuH}
\eeq
by comparing to the theory prediction in Ref.~\cite{Antognini:2012ofa} that relies on a data-driven evaluation of the proton polarizability contribution \cite{Carlson:2011af}:  
\begin{equation}
E_\mathrm{hfs}^\mathrm{pol.}(2S,\mu\text{H})=8.0(2.6)\, \upmu\mathrm{eV}.
\end{equation}
Several collaborations are now preparing a measurement of the ground-state ($1S$) hfs in $\mu$H with ppm precision:
CREMA \cite{Amaro:2021goz}, FAMU \cite{Pizzolotto:2021dai, Pizzolotto:2020fue} and J-PARC \cite{Sato:2014uza}  (see Ref.~\cite{Pohl:2016xsr} for a comparison of the different experimental methods). These future measurements hold the potential to extract the Zemach radius with a sub-percent uncertainty, thereby constraining the magnetic properties of the proton.

A precise theory prediction for the $1S$ hfs in $\mu$H is essential for the success of the experimental campaigns. Firstly, to narrow down the frequency search range, which is important given the limited beam time available to the collaborations at PSI, RIKEN-RAL and J-PARC. Secondly, for the interpretation of the results. One can either extract the Zemach radius given a theory prediction for the proton-polarizability effect in the $\mu$H $1S$ hfs, or vice versa, extract the proton-polarizability effect with input for the Zemach radius. Furthermore, one can combine the precise measurements of the $1S$ hfs in H and $\mu$H to disentangle the Zemach radius and polarizability effects, leveraging radiative corrections as explained in Ref.~\cite{Antognini:2022xoo}, and compare their empirical values to theoretical expectations.

The biggest uncertainty in the theory prediction comes from proton-structure effects, entering through the two-photon exchange (TPE). These contain the above-mentioned Zemach radius and polarizability effects. Presently, they are evaluated within a ``data-driven'' dispersive approach \cite{Carlson:2008ke,Faustov:2006ve,Tomalak:2018uhr}. While the dispersive method itself is rigorous, it requires sufficient experimental data to map out the proton spin structure functions $g_1(x,Q^2)$ and $g_2(x,Q^2)$ as full functions of the Bjorken variable $x$ and the photon virtuality $Q^2$. This has been the aim of a dedicated ``Spin Physics Program'' at Jefferson Lab \cite{Chen:2008ng,CLAS:2017ozc,CLAS:2021apd,JeffersonLabE97-110:2019fsc,E97-110:2021mxm} that recently extended the previously scarce data for  $g_2$ \cite{Zielinski:2017gwp,JeffersonLabHallAg2p:2022qap}.

In this work, we use an entirely different approach
--- the chiral perturbation theory ($\chi$PT) \cite{Weinberg:1978kz,Gasser:1983yg,Gasser:1987rb} --- which has been successfully used  to give a prediction for the proton-polarizability effect in the $\mu$H Lamb shift \cite{Alarcon:2013cba}. To be precise, we work in the framework of baryon chiral perturbation theory (B$\chi$PT) --- the manifestly Lorentz-invariant formulation of $\chi$PT in the baryon sector~\cite{Gasser:1987rb, Gegelia:1999qt,Fuchs:2003qc} (see also \cite{Pascalutsa:2006up,Geng:2013xn} for reviews). We show that the leading-order (LO) B$\chi$PT prediction for the polarizability effect in the hfs is effectively vanishing,  thereby, in substantial disagreement with the data-driven evaluations.

The paper is organized as follows. In \secref{Calculation}, we discuss the forward TPE, and in particular, the polarizability effect in the hfs. A new formalism where one splits into contributions from the longitudinal-transverse and helicity-difference photoabsorption cross sections of the proton, $\sigma_{LT}$ and $\sigma_{TT}$, is introduced in \Eqref{POLAlternative}. It will be shown that this decomposition is advantageous for both the dispersive, as well as the effective field theory (EFT) calculations, as it gives a cleaner access to the uncertainties. More details are given in Appendix \ref{AppendixTPE}.  In \secref{PionCloud}, we present our LO  B$\chi$PT prediction for the polarizability effect in the hfs of H and $\mu$H, together with a detailed discussion of the uncertainty estimate. In \secref{Comparison}, we compare our results to data-driven dispersive and heavy baryon effective field theory (HB EFT) calculations.  In \secref{ZemachRadius}, the Zemach radius is extracted from H and $\mu$H spectroscopy based on our prediction for the polarizability effect. In \secref{TPE}, we discuss the TPE effect in the $\mu$H hfs in view of the forthcoming experiments. Full details of the theoretical prediction for the $1S$ $\mu$H hfs are collected in Appendix \ref{TheorySummary}. We finish with an outlook and conclusions.

\section{Two-photon exchange in the hyperfine splitting}
\seclab{Calculation}

\begin{figure}[t]
\centering
\begin{minipage}{0.3\textwidth}
\centering
  \includegraphics[width=6cm]{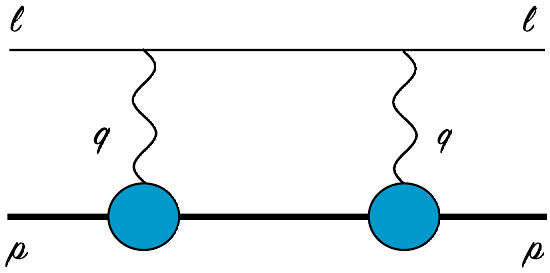}
  
  (a)
\end{minipage}
\hspace{3cm}
\begin{minipage}{0.3\textwidth}
\centering
       \includegraphics[width=6cm]{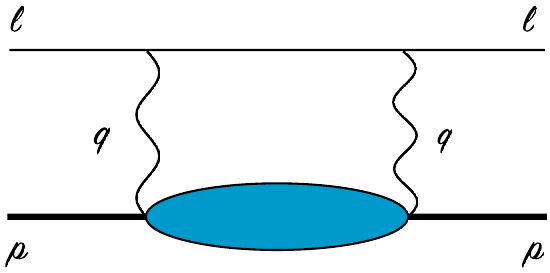}

      (b)
    \end{minipage}
\caption{Two-photon-exchange diagram in forward kinematics: (a) Elastic contribution; (b) polarizability contribution. The horizontal lines correspond to the lepton and the proton (bold), where the `blob' represents all possible excitations. The crossed diagrams are not drawn.\label{fig:TPE}}
\end{figure}

The (muonic-)hydrogen hfs receives contributions from QED-, weak- and strong-interaction effects:
 \beq
E_{\mathrm{hfs}}(nS)=\frac{E_\mathrm{F}}{n^3}\left(1+\Delta_\mathrm{QED}+\Delta_\mathrm{weak}+\Delta_\mathrm{strong}\right),\eqlab{hfswoEF}
\eeq
where the leading-order in $\alpha$ contribution is given by the  Fermi energy:
\beq
E_\mathrm{F}=\frac{8Z\al}{3a^3}\frac{1+\kappa}{mM},\eqlab{FermiE}
\eeq
with $\al$ the fine-structure constant, $Z$ the charge of the nucleus (in the following $Z=1$ for the proton), $m$, $M$ the lepton and proton masses, $\kappa$ the anomalous magnetic moment of the proton, and $a^{-1}=\al m_r$ the inverse Bohr radius, with $m_r=mM/(m+M)$ the reduced mass. The strong-interaction effects arise from the composite structure of the proton. They begin to enter at $\mathcal{O}(\al^5)$, see for instance Ref.~\cite{Carlson:2008ke}, where they are split into the Zemach-radius, recoil, and polarizability contributions:
\beq
\Delta_\mathrm{strong}=\Delta_\mathrm{Z}+\Delta_\mathrm{recoil}+\Delta_\mathrm{pol.}\,,\eqlab{FS_hfs}
\eeq
which can all be attributed to the forward
TPE shown in \Figref{TPE}. For a first comprehensive theory summary of the Lamb shift, fine and hyperfine structure in $\mu$H, including proton-structure dependent effects, we refer to Ref.~\cite{Pachucki:1996zza}. The Zemach and recoil terms ($\Delta_\mathrm{Z}$ and  $\Delta_\mathrm{recoil}$) are elastic contributions with a proton in the intermediate state, see \Figref{TPE} (a). The diagram in \Figref{TPE} (b) contains excited intermediate
states ($\pi N$, $\De$-isobar, etc.) represented by the `blob'. It generates the polarizability effect ($\Delta_\mathrm{pol.}$) that shall be evaluated in this work. 

The forward TPE contribution to the hfs can be expressed through the spin-dependent forward doubly-virtual Compton scattering (VVCS) amplitudes, $S_1$ and $S_2$, cf.\ \Eqref{VVCS_TA}. The latter can be related to the proton structure functions $g_1$ and $g_2$ in a dispersive approach, cf.\ Eqs.~\eref{LTTT} and \eref{S12nB}.
A full derivation of the well-known formalism for the TPE contribution to the hfs can be found in Appendix \ref{AppendixTPE}.

The largest TPE effect is due to the Zemach radius contribution: 
\beq
\Delta_\mathrm{Z}=-2\al m_r R_\mathrm{Z}.\eqlab{ZemachTerm} 
\eeq
The recoil contribution is one order of magnitude smaller \cite{Carlson:2011af}, and will not be considered in this paper. It has been recently updated in Ref.~\cite{Antognini:2022xqf}.
The hfs is therefore best suited for a precision extraction of the Zemach radius, defined as the following integral over the electric and magnetic Sachs form factors $G_{E}(Q^2)$ and $G_{M}(Q^2)$ \cite{Pachucki:1996zza}: 
\beq
\eqlab{RZdef}
R_\mathrm{Z}=-\frac{4}{\pi}\int_0^\infty \frac{\dd Q}{Q^2} \left[\frac{G_{E}(Q^2)G_{M}(Q^2)}{1+\kappa}-1\right],
\eeq
where $q^2=-Q^2$ is the photon virtuality. Equivalently, we can write:
\beq
\eqlab{RZdef2}
R_\mathrm{Z}=\langle r\rangle_E+\langle r\rangle_M-\frac{2}{\pi^2} \int_0^\infty \frac{\mathrm{d}t}{t}\frac{\im G_M(t)}{1+\kappa}\int_0^\infty \frac{\mathrm{d}t'}{t'}\frac{\im G_E(t')}{\sqrt{t}+\sqrt{t'}},
\eeq
where the linear electric and magnetic radii are defined as:
\beq
\langle r\rangle_{E,M}=\frac{2}{\pi} \int_0^\infty \frac{\mathrm{d}t}{t^{3/2}}\im G(t),
\eeq
with $\im G(t)$ the imaginary part of the normalized electric or magnetic Sachs form factor, $G_{E,M}(Q^2)/G_{E,M}(0)$. As one can see from Eqs.~\eref{RZdef} and \eref{RZdef2}, a measurement of the Zemach radius gives access to the magnetic properties of the proton.

The polarizability effect in the hfs is fully constrained by empirical information on the proton spin structure functions $g_1(x,Q^2)$ and $g_2(x,Q^2)$,
and the Pauli form factor $F_2(Q^2)$, functions of $Q^2$ and the Bjorken variable $x=Q^2/2M\nu$, where $\nu$ is the photon energy in the lab frame. This is in contrast to the Lamb shift, where the knowledge of a subtraction function, $T_1(0,Q^2)$ or $T_1(iQ,Q^2)$ \cite{Hagelstein:2020awq}, is needed.\footnote{See Ref.~\cite{Biloshytskyi:2023fyv} for a recent proposal how the subtraction functions can be related to integrals over photoabsorption cross sections.}
It reads:\footnote{Note that our notation largely follows Ref.~\cite{Hagelstein:2015egb}. It differs slightly from other literature, where $\delta_i$ is usually denoted $\Delta_i$ \cite{Nazaryan:2005zc}.}
\begin{subequations}
\eqlab{POL}
\begin{align}
\Delta_\mathrm{pol.}&=\Delta_1+\Delta_2=\frac{\al m}{2\pi (1+\kappa) M}\big( \delta_1+\delta_2\big),\\
\delta_1&=2\int_0^\infty\frac{\dd Q}{Q}\left\{\frac{5+4v_l}{(v_l+1)^2}\Big[4I_1(Q^2)+F_{2}^2(Q^2)\Big]-\frac{32M^4}{Q^4}\int_0^{x_0}\dd x\, x^2g_1(x,Q^2)\right.\eqlab{Delta1b}\\
&  \times \left. \frac{1}{(v_l+ v_x)(1+ v_x)(1+v_l)}\left(4+\frac{1}{1+ v_x}+\frac{1}{v_l+1}\right) \right\}
,\nn\qquad\\
\delta_2&=96M^2\int_0^\infty\frac{\dd Q}{Q^3}\int_0^{x_0}\dd x\,g_2(x,Q^2) \left(\frac{1}{v_l+ v_x}-\frac{1}{v_l+1} \right),\eqlab{Delta2}
\end{align}
\end{subequations}
with $x_0$ the inelastic threshold, $v_l=\sqrt{1+\nicefrac{1}{\tau_l}}$, $v_x=\sqrt{1+x^2\tau^{-1}}$, $\tau_l=\nicefrac{Q^2}{4m^2}$, $\tau=\nicefrac{Q^2}{4M^2}$, and the generalized Gerasimov-Drell-Hearn (GDH) integral: 
\beq
\eqlab{I1def}
I_1(Q^2)=\frac{2M^2}{Q^2}\int_0^{x_0}\dd x\, g_1(x,Q^2)=\bar I_1(Q^2)-F_{2}^2(Q^2)/4.
\eeq
Here, $\bar I_1$ is the polarizability part of $I_1$. For the origin of the Pauli form factor in the above equations, see  discussion in Appendix \ref{AppendixTPE}.

As we will show in \secref{Sec3B}, instead of decomposing into $\Delta_1$ and $\Delta_2$, it is convenient to decompose into contributions from the longitudinal-transverse and helicity-difference cross sections $\sigma_{LT}$ and $\sigma_{TT}$: 
\begin{subequations}
\eqlab{POLAlternative}
\beq
\Delta_\mathrm{pol.}=\Delta_{LT}+\Delta_{TT}+\Delta_{F_2}=\frac{\al m}{2\pi (1+\kappa) M}\left(\delta_{LT}+\delta_{TT}+\delta_{F_2}\right),
\eeq
where we define:
\begin{align}
&\delta_{LT}=\frac{4M}{\al\pi^2}\int_0^\infty \!\dd Q\int_0^{x_0}\!\dd x\,\frac{1}{v_l+v_x}\frac{1}{x^2+\tau}\left[1-\frac{1}{(1+v_l)(1+v_x)}\right]\sigma_{LT}(x,Q^2)\eqlab{DeltaLT},\\
&\delta_{TT}=\frac{4M^2}{\al\pi^2}\int_0^\infty \frac{\dd Q}{Q}\int_0^{x_0}\frac{\dd x}{x} \frac{1}{1+v_l}\left[\frac{2\tau}{x^2+\tau}+\frac{1}{(v_l+v_x)(1+v_x)}\right]\si_{TT}(x,Q^2),\eqlab{DeltaTT}\\
&\delta_{F_2}=2\int_0^\infty\frac{\dd Q}{Q}\frac{5+4v_l}{(v_l+1)^2}\,F_{2}^2(Q^2).
\end{align}
\end{subequations} 
Or equivalently, in terms of the VVCS amplitudes, we can write:
\begin{subequations}
\eqlab{POLAlternative2}
\begin{align}
\delta_{LT}&=\frac{8M}{\al}\frac{1}{(2\pi)^3} \frac{1}{i}\int_{-\infty}^\infty \!\dd\nu\!\int\!\dd \bq\,\frac{1}{Q^4-4m^2\nu^2}\left\{\bar S_1(\nu,Q^2)+\frac{\nu}{M}\bar S_2(\nu,Q^2)\right\}\nn\\
&-\frac{2}{m^2}\int_0^\infty\dd Q\,Q\,(v_l-1)\,F_{2}^2(Q^2),\\
\delta_{TT}&=\frac{4M}{\al}\frac{1}{(2\pi)^3} \frac{1}{i}\int_{-\infty}^\infty \!\dd\nu\!\int\!\dd \bq\,\frac{1}{Q^4-4m^2\nu^2}\left\{\frac{\nu}{M}\bar S_2(\nu,Q^2)-\frac{\nu^2}{Q^2} \bar S_1(\nu,Q^2)\right\}\nn\\
&-2\int_0^\infty\frac{\dd Q}{Q}\frac{1}{(v_l+1)^2}\,F_{2}^2(Q^2).
\end{align}
\end{subequations} 
Here, $\bar S_i$ denotes the non-Born part of the amplitudes.
An advantage of the B$\chi$PT calculation in this work is that the non-Born amplitudes can be calculated directly, and need not be constructed through the dispersive formalism. Furthermore, at the present order of our calculation in the B$\chi$PT power counting, there are no contributions to the elastic form factors, and thus, $I_1$ in \Eqref{I1def} is given by the polarizability part only. 

\section{Chiral loops} \seclab{PionCloud}

Assuming B$\chi$PT is an adequate theory of low-energy nucleon structure, it should be well applicable to atomic systems, where the relevant energies are naturally small. In Ref.~\cite{Alarcon:2013cba}, the polarizability effect in the $\mu$H Lamb shift has been successfully predicted at  LO in B$\chi$PT. Here, we extend this calculation to the polarizability effect in the hfs. This requires the spin-dependent non-Born VVCS amplitudes, $\bar S_1$ and $\bar S_2$,  at chiral $\mathcal{O}(p^3)$ in the B$\chi$PT power counting. 
 
Figure 1 in Ref.~\cite{Alarcon:2013cba} shows the leading polarizability effect given by the TPE diagrams of elastic lepton-proton scattering with one-loop $\pi N$ insertions. For the Compton-like processes, it is convenient to use 
the chirally-rotated leading B$\chi$PT Lagrangian  for the pion $\pi^a(x)$ and nucleon $N(x)$ fields \cite{Len10}:
\bea
\mathcal{L}_{\pi N}^{(1)}&=&\bar{N} \Big( i \slashed{\partial}-M_N-i \frac{g_A}{f_\pi}M_N \tau^a\pi^a\gamma_5 +\frac{g_A^2}{2f_\pi^2}M_N \pi^2+\frac{g_A^2-1}{4f_\pi^2}\tau^a \epsilon^{abc}\pi^b \slashed{\partial}\,\pi^c\Big) N+\mathcal{O}(\pi^3),
\eea
where $\ga_5=i \ga^0 \ga^1 \ga^2 \ga^3$, $g_A\simeq 1.27$ \cite{Olive:2016xmw} is the axial coupling of the nucleon, $f_\pi \simeq 92.21 $ MeV is the pion-decay constant, $\tau^a$ are the Pauli matrices, $M_N\simeq  938.27$ MeV and $m_\pi\simeq 139.57$ MeV are the nucleon and pion masses.\footnote{Note that isospin-breaking effects, such as differences in nucleon or pion masses, are neglected in the loops.} 
As described in Ref.~\cite{Alarcon:2013cba}, the Born part is separated from the $\mathcal{O}(p^3)$ VVCS amplitudes by subtracting the on-shell pion-loop $\gamma N N$-vertex in the one-particle-reducible VVCS graphs, see diagrams (b) and (c) in Figure 1 of Ref.~\cite{Alarcon:2013cba}. For more details on the B$\chi$PT framework, we refer to Refs.~\cite{Lensky:2014dda,Alarcon:2020wjg,Alarcon:2020icz}, where the complete next-to-next-to-leading-order (NNLO) in the $\delta$-expansion \cite{Pascalutsa:2003aa} B$\chi$PT calculation of
the spin-independent and spin-dependent nucleon VVCS amplitudes can be found.\footnote{See also Refs.~\cite{Bernard:2002pw, Bernard200882,Bernard:2012hb} for nucleon VVCS studies in B$\chi$PT within the  $\epsilon$-expansion power-counting scheme \cite{Hemmert:1996xg}.} 

In practice, most results here were obtained based on our B$\chi$PT prediction for the $\pi N$-production channel in the structure functions $g_i$, given in Ref.~\cite[Appendix B]{Alarcon:2020icz}. It has been verified that the results agree with the calculation based on the VVCS amplitudes $\bar S_i$.

\subsection{Numerical results}\seclab{HBExpPion}

Our LO B$\chi$PT prediction for the polarizability effect in the $1S$ hfs of H and $\mu$H amounts to:
\begin{subequations}
\eqlab{pionresulthfs}
\bea
E_\mathrm{hfs}^{{\langle \mathrm{LO}\rangle}\,\mathrm{pol.}}(1S, \mathrm{H})&=&0.69
(2.03)\,\mathrm{peV},\eqlab{pionresulthfs2SH}\\
E_\mathrm{hfs}^{{\langle \mathrm{LO}\rangle}\,\mathrm{pol.}}(1S, \mu\mathrm{H})&=&6.8(11.4)\,\upmu\mathrm{eV}.\eqlab{pionresulthfs2SmuH}
\eea
\end{subequations}
The error estimate will be described and motivated in the subsequent sections.
The corresponding contributions to the $nS$ hfs are trivially obtained through a $1/n^3$ scaling, as can be seen from Eqs.~\eref{hfswoEF} and \eref{FermiE}. Splitting into contributions from the spin structure functions $g_1$ and $g_2$, we obtain:
\begin{subequations}
\bea
E_\mathrm{hfs}^{{\langle \mathrm{LO}\rangle}\,\mathrm{pol.}}(1S, \mathrm{H}, \Delta_1)&=&0.3(3.1)\,\mathrm{peV},\qquad
E_\mathrm{hfs}^{{\langle \mathrm{LO}\rangle}\,\mathrm{pol.}}(1S, \mathrm{H}, \Delta_2)=0.4(1.0)\,\mathrm{peV},\\
E_\mathrm{hfs}^{{\langle \mathrm{LO}\rangle}\,\mathrm{pol.}}(1S, \mu\mathrm{H}, \Delta_1)&=&5.2(16.5)\,\upmu\mathrm{eV},\qquad E_\mathrm{hfs}^{{\langle \mathrm{LO}\rangle}\,\mathrm{pol.}}(1S, \mu\mathrm{H},\Delta_2)=1.6(5.2)\,\upmu\mathrm{eV}.
\eea
Strikingly, the contributions from the longitudinal-transverse and helicity-difference cross sections $\sigma_{LT}$ and $\sigma_{TT}$:  
\bea
E_\mathrm{hfs}^{{\langle \mathrm{LO}\rangle}\,\mathrm{pol.}}(1S, \mathrm{H}, \Delta_{LT})&=&5.1(1.5)\,\mathrm{peV},\qquad
E_\mathrm{hfs}^{{\langle \mathrm{LO}\rangle}\,\mathrm{pol.}}(1S, \mathrm{H}, \Delta_{TT})=-4.4(1.3)\,\mathrm{peV},\\
E_\mathrm{hfs}^{{\langle \mathrm{LO}\rangle}\,\mathrm{pol.}}(1S, \mu\mathrm{H}, \Delta_{LT})&=&30.0(9.0)\,\upmu\mathrm{eV},\qquad E_\mathrm{hfs}^{{\langle \mathrm{LO}\rangle}\,\mathrm{pol.}}(1S, \mu\mathrm{H},\Delta_{TT})=-23.2(7.0)\,\upmu\mathrm{eV},
\eea
\end{subequations}
are one order of magnitude larger than the total, and differ in their respective signs. This indicates a cancellation of LO contributions between $\Delta_{LT}$ and $\Delta_{TT}$.

Including in addition the correction due to  electron vacuum polarization (eVP) in the TPE diagram, see \Figref{TPEwithVP} and discussion in Appendix \ref{eVPapp}, gives a negligible effect within the present uncertainties:
\begin{subequations}
\eqlab{eVPnumbers}
\bea
E_\mathrm{hfs}^{{\langle \mathrm{LO}\rangle}\,\text{pol. + eVP}}(1S, \mathrm{H})&=&0.72 (2.07)\,\mathrm{peV},\\
E_\mathrm{hfs}^{{\langle \mathrm{LO}\rangle}\,\text{pol. + eVP}}(1S, \mu\mathrm{H})&=&7.0(11.6)\,\upmu\mathrm{eV}.\eqlab{eVPnumbersmuH}
\eea
\end{subequations}
Nevertheless, it is important in view of the anticipated $1$ ppm accuracy (corresponding to $\sim 0.2\,\upmu$eV) of the $\mu$H $1S$ hfs measurement by the CREMA collaboration \cite{Amaro:2021goz}. We therefore include the additional $\Delta_\mathrm{pol.}^\mathrm{eVP} (\text{H})=0.01$ ppm and $\Delta_\mathrm{pol.}^\mathrm{eVP}(\mu\text{H})=1$ ppm on top of $\Delta_\mathrm{pol.} (\text{H})=0.12(35)$ ppm and $\Delta_\mathrm{pol.}(\mu\text{H})=37(62)$ ppm.

To understand why the contributions from $\sigma_{LT}$ and $\sigma_{TT}$ largely cancel in $\Delta_\mathrm{pol.}$, we study the heavy-baryon (HB) limit of the spin-dependent VVCS amplitudes \cite{Ji:1999mr}. Expanding the LO B$\chi$PT expression for the $\bar S_1$ amplitude in $\mu=m_\pi/M_N$ while keeping the ratio of the light scales $\tau_\pi =\nicefrac{Q^2}{4m_\pi^2}$ fixed, one obtains:
\beq
\bar S_1(0,Q^2)\!\stackrel{\text{HB}}{=}\!-\frac{3\,\al\, g_A^2}{16 f_\pi^2} \,m_\pi\!\left[1-(1+\tau_\pi)\frac{\arctan\sqrt{\tau_\pi}}{\sqrt{\tau_\pi}}\right]\!.\eqlab{28a}
\eeq
We then take a closer look at the first term in the low-energy polarizability expansion:
\beq
\eqlab{S10Q2polexp}
\frac{\bar S_1(0,Q^2)}{Q^2}\Big\vert_{Q^2\rightarrow 0}=M_N\left\{\gamma_{E1M2}-3\al M_N \left[P^{\prime(M1,M1)1}(0)+P^{\prime(L1,L1)1}(0)\right]\right\}.
\eeq
The HB$\chi$PT predictions for the  proton polarizabilities \cite{Kum00,Hemmert:1996gr,Hemmert:1997at,Hemmert:1999pz,Kao:2002cn,Kao:2004us} entering \Eqref{S10Q2polexp} read:
\begin{subequations}
\eqlab{individualPol}
\bea
\gamma_{E1M2}&=&\frac{\al g_A^2}{(4\pi f_\pi)^2}\frac{1}{m_\pi^2}\frac{1}{6}\left[1-\frac{7\pi}{4}\frac{m_\pi}{M_N}\right]\!,\\
P^{\prime(M1,M1)1}(0)&=&\frac{g_A^2}{(4\pi f_\pi)^2}\frac{1}{m_\pi^2}\frac{1}{18M}\left[-1+\frac{7\pi}{4}\frac{m_\pi}{M_N}\right]\!,\\
P^{\prime(L1,L1)1}(0)&=&\frac{g_A^2}{(4\pi f_\pi)^2}\frac{1}{m_\pi^2}\frac{1}{9M}\left[1-\frac{17\pi}{8}\frac{m_\pi}{M_N}\right]\!.
\eea
\end{subequations}
We can see that the leading terms in the chiral expansion are of $\mathcal{O}(1/m_\pi^2)$. They cancel among the different polarizabilities, thus,  \Eqref{HBS1} becomes a subleading contribution:
\beq
\frac{\bar S_1(0,Q^2)}{Q^2}\Big\vert_{Q^2\rightarrow 0}\stackrel{\text{HB}}{=}\frac{\alpha  g_A^2}{32 f_\pi^2}\frac{1}{m_\pi}.\eqlab{HBS1}
\eeq
Accordingly, one would expect the chiral loops in the hfs to be small. Indeed, the LO B$\chi$PT prediction in \Eqref{pionresulthfs} is essentially vanishing, where the small number is mainly a remnant of higher orders in the HB expansion.
This has to be taken into account in the uncertainty estimate.

\begin{figure}[t]
\centering
\includegraphics[width=0.45\columnwidth]{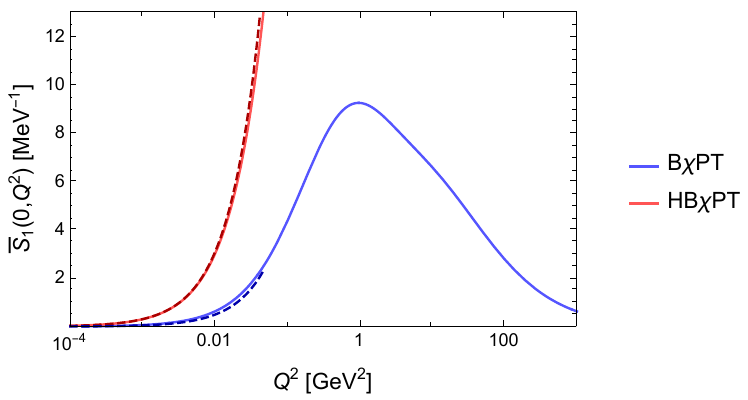}
\caption{The amplitude $\bar S_1(0,Q^2)$ at LO in B$\chi$PT (blue) and HB$\chi$PT (red). The dashed lines show the corresponding
slope terms, i.e., the first terms in the expansion in powers of $Q^2$. The B$\chi$PT slope has been calculated from the polarizabilities given in Ref.~\cite[Table I]{Lensky:2017dlc}, the HB$\chi$PT slope is given in \Eqref{HBS1}.
}
\label{S1fig}
\end{figure}

Note that the HB expansion above has been introduced  for instructive purposes only, but is not entering our calculation of the polarizability effect. 
The HB$\chi$PT prediction of the $S_1(0,Q^2)$ amplitude, Eq.~\eref{28a}, raises with $Q$,
thus, its contribution to the hfs will be divergent. This can be seen from Fig.~\ref{S1fig}, where we compare the chiral-loop contribution to $\bar S_1(0,Q^2)$ as predicted by B$\chi$PT and HB$\chi$PT, respectively.

\subsection{Uncertainty estimate}
\seclab{Sec3B}
 
\begin{figure}[tbh]
\centering
\begin{minipage}{0.49\textwidth}
\centering
  \includegraphics[scale=0.48]{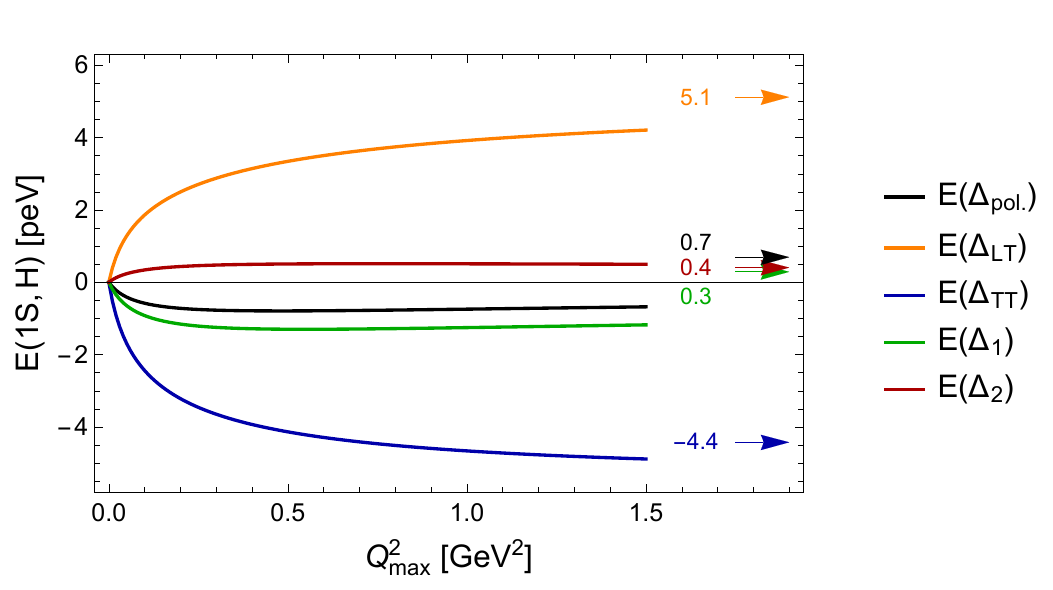}
\end{minipage}
\hfill
\begin{minipage}{0.49\textwidth}
\centering
       \includegraphics[scale=0.48]{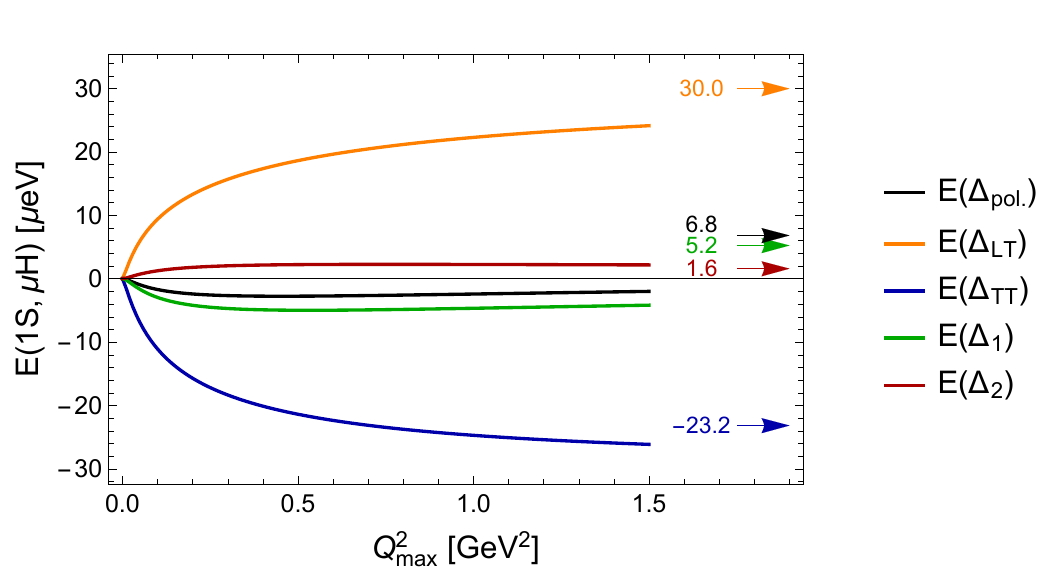}
    \end{minipage}
\caption{Polarizability effect on the $1S$ hyperfine splitting in H (left panel) and $\mu$H (right panel): Cutoff dependence of the leading-order $\pi N$-loop contribution. The total results, Eqs.~\eref{pionresulthfs2SH} and \eref{pionresulthfs2SmuH}, are indicated by the black arrows.}
\label{fig1piPeH}
\end{figure}

B$\chi$PT is a low-energy EFT of QCD describing strong interactions in terms of hadronic degrees of freedom (pion, nucleon, $\Delta(1232)$ resonance). An important requirement for a reliable B$\chi$PT prediction is that the contribution from beyond the scale at which this EFT is safely applicable, i.e., $Q_\mathrm{max}>m_\rho=775$ MeV, has to be small. For the LO B$\chi$PT prediction of the polarizability effect in the $\mu$H Lamb shift \cite{Alarcon:2013cba}, the contribution from beyond this scale was less than $15\,\%$, thus, within the expected uncertainty.  Comparing the TPE master formulas for Lamb shift and hfs, Eqs.~\eref{LSMaster} and \eref{VVCS_hfs}, the weighting function in the former has a stronger suppression for large $Q^2$. It is therefore important to verify that the same quality criterion still holds for the hfs prediction presented here.

Let us consider the polarizability effect as a running integral with momentum cutoff $Q_\mathrm{max}$, as shown in Fig.~\ref{fig1piPeH}. The convergence of the $\Delta_1$ (green line) contribution, as well as of the total $\Delta_\mathrm{pol.}$ (black line), is poor. They display a sign change of the running integral at energies above $Q_\mathrm{max}\approx 2$ GeV ($\mu\mathrm{H}$) and $\approx 4$ GeV ($\mathrm{H}$), respectively.
$\Delta_2$ (red line) converges better. Its contributions from above $Q_\mathrm{max}=m_\rho$ amount to $42\,\%$ (H) and $26\,\%$ ($\mu\mathrm{H}$), respectively.

The bad high-momentum asymptotics indicated above are merely an artefact of the conventional splitting into $\Delta_1$ and $\Delta_2$. For the alternative splitting into $\Delta_{LT}$ and $\Delta_{TT}$, introduced in \Eqref{POLAlternative}, the cut-off dependence improves considerably. For $\Delta_{TT}$ (blue line), the contribution from above $Q_\mathrm{max}=m_\rho$ amounts to less than $4\,\%$ for both hydrogens. For $\Delta_{LT}$ (orange line), the high-energy contributions are less than $35\,\%$ ($\mu\mathrm{H}$) and $32\,\%$ ($\mathrm{H}$), respectively. In this way, our results are in agreement with the natural expectation of uncertainty for a LO prediction, $30\,\%$ [$\simeq (M_\Delta-M)/\mathrm{GeV}$], in B$\chi$PT with inclusion of the $\Delta$ resonance.
Based on this analysis, we decided to assign errors of $30\,\%$  to the $\si_{LT}$ and $\si_{TT}$ contributions, and propagate them to $\Delta_1$, $\Delta_2$ and $\Delta_\mathrm{pol.}$. It is interesting to note that in this way the uncertainty of $\Delta_1$ is larger than the uncertainty of $\Delta_\mathrm{pol.}$. This can be understood from the opposite signs of the $\Delta_{i,j}$ contributions, where $i=1,2$ and $j=LT,TT$, on the example of $\mu$H: 
\begin{subequations}
\bea
\Delta_{1,\,LT}=&227\,\text{ppm},\qquad \Delta_{1,\,TT}&=-198\,\text{ppm},\\ \qquad \Delta_{2,\,LT}=&-62\,\text{ppm},\qquad \Delta_{2,\,TT}&=71\,\text{ppm}.
\eea
\end{subequations}

\section{Comparison with other results} \seclab{Comparison}

In this section, we compare our LO B$\chi$PT prediction for the polarizability effect in the $\mathrm{H}$ and $\mu\mathrm{H}$ hfs to other available evaluations. Furthermore, we study the contribution of the $\bar S_1(0,Q^2)$ subtraction function and the scaling of the polarizability effect with the lepton mass.

\subsection{Heavy-baryon effective field theory}\label{ChPTComp}

Let us start by comparing our B$\chi$PT prediction to other model-independent calculations using HB EFT \cite{Pineda:2003,Peset:2014jxa,Peset:2016wjq}.\footnote{Full details on the B$\chi$PT framework used in here, and how it distinguishes from HB EFT,  can be found in Refs.~\cite{Alarcon:2013cba}.} First results for the elastic and inelastic TPE effects on the hfs in H and $\mu\mathrm{H}$ have been obtained in Ref.~\cite{Pineda:2003}, where the contribution of the leading chiral logarithms, $\mathcal{O}(m^3 \al^5/M^2\times\left[\ln m_\pi, \ln \Delta, \ln m\right])$, was calculated in HB EFT matched to potential NRQED. At this order in the chiral expansion the polarizability effects in the hfs from pion-nucleon and pion-delta loops cancel each other in the large-$N_c$ limit, while the $\Delta$ exchange cancels part of the point-like corrections, see also Ref.~\cite{Ji:1999mr}. The analytical results presented in Ref.~\cite{Pineda:2003,Pineda:2003hb} motivate the relative size of the Zemach and polarizability corrections.  

Updated HB EFT predictions for the TPE effects on the hydrogen spectra can be found in Refs.~\cite{Peset:2014jxa,Peset:2014yha,Peset:2016wjq}. In Ref.~\cite{Peset:2016wjq}, the difference between the pion-loop polarizability contributions in $\mathrm{H}$ and $\mu\mathrm{H}$ is quoted as
\beq
\Delta c_4\equiv c_{4,\mathrm{pol}}^{\mu\mathrm{H}}-c_{4,\mathrm{pol}}^{\mathrm{H}}=0.17(9), \;
\eeq
where $c_4$ is a Wilson coefficient linked to the hfs in the following way:
\beq
E_\mathrm{hfs}(nS)=\frac{E_\mathrm{F}}{n^3}\frac{3\al}{2 \pi(1+\kappa)}\frac{m}{M}\,c_4.
\eeq
For comparison, we can evaluate the analogue of $\Delta c_4$ from other theory predictions for the polarizability contribution.
Within errors, our LO B$\chi$PT prediction agrees with this result:
\beq
\Delta c_4=
0.09(0.46).
\eeq
Here, the uncertainties of the H and $\mu$H predictions have been combined in quadrature to estimate the error on their difference.
For comparison, from the data-driven dispersive evaluations of Carlson et al.\ \cite{Carlson:2008ke}, one can deduce:
\beq
\Delta c_4=-0.27(1.53),
\eeq
where we combined all errors quoted in Ref.~\cite{Carlson:2008ke} and estimated the error on $\Delta c_4$ in the same way as done above.

\subsection{Data-driven dispersive evaluations}\label{DataComp}

\begin{figure}[t]
\centering
\includegraphics[width=0.6\columnwidth]{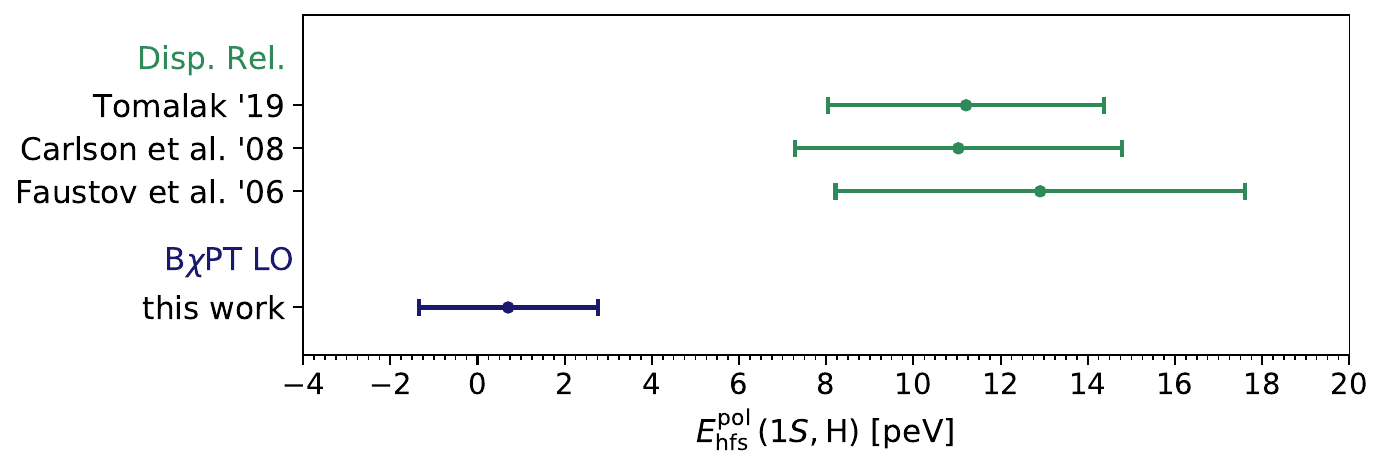}
\includegraphics[width=0.6\columnwidth]{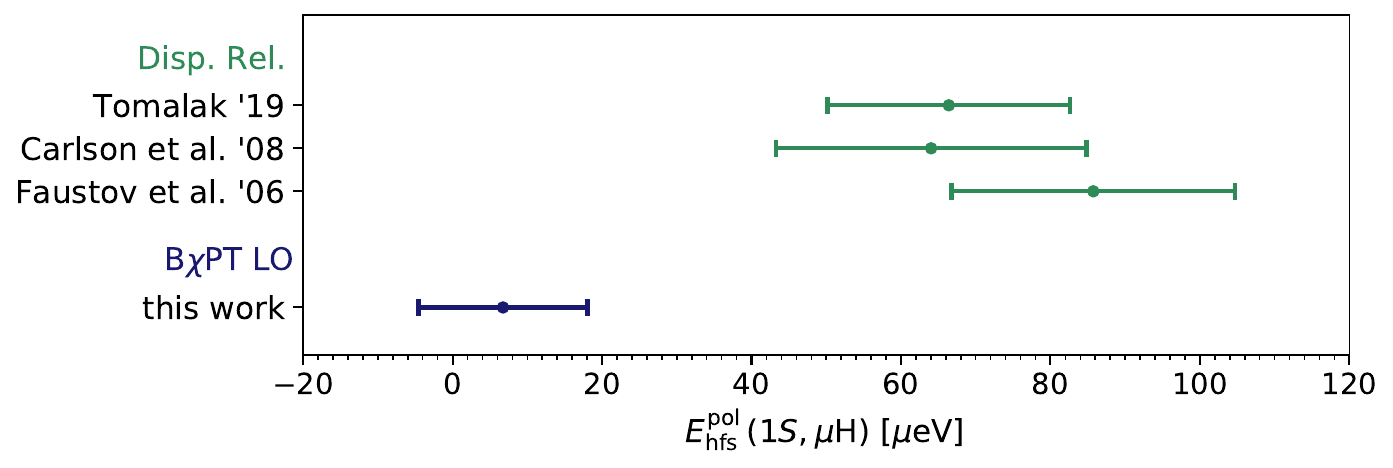}
\caption{Comparison of available results for the  polarizability effect on the hyperfine splitting in H and $\mu$H (upper and lower panel) \cite{Carlson:2008ke,Faustov:2006ve,Tomalak:2018uhr}.}
\label{fig:CompEH}
\end{figure}

There is a clear discrepancy between the B$\chi$PT prediction, presented here, and the conventional data-driven dispersive evaluations. The dispersive evaluations rely on empirical information for the inelastic proton spin structure functions, the elastic Pauli form factor and polarizabilities.
The discrepancy can be seen from Fig.~\ref{fig:CompEH}, where our LO B$\chi$PT prediction for the polarizability effect in the $\mathrm{H}$ and $\mu\mathrm{H}$ hfs is compared to the available dispersive evaluations.  Adding an estimate for the next-to-leading-order (NLO) effect of the $\Delta(1232)$ resonance \cite{Hagelstein:2018bdi}, obtained from large-$N_c$ relations  for the nucleon-to-delta transition form factors, to the model-independent LO B$\chi$PT prediction will improve agreement for $\Delta_2$ but not for $\Delta_\mathrm{pol}$.

The origin of this discrepancy has to be understood in order to give a reliable prediction of the TPE effect in the $\mu$H hfs, needed for the forthcoming experiments. Part of the discrepancy might be due to underestimated uncertainties. An evaluation of the total polarizability  effect suffers from cancellations in two places: firstly, between contributions from the cross sections $\sigma_{LT}$ and $\sigma_{TT}$, secondly, between the elastic Pauli form factor $F_2$ and the inelastic structure functions in the low-$Q$ region. Each of these cancellations reduces the result by an order of magnitude. In the calculation presented here, the former is taken into account by estimating the uncertainty due to higher-order corrections in the B$\chi$PT power counting based on the large $\sigma_{LT}$ and $\sigma_{TT}$ contributions, see discussion in \secref{Sec3B}. In the dispersive approach, it would be important to take into account correlations between parametrizations of the $g_1$ and $g_2$ structure functions, which both rely on measurements of $\sigma_{LT}$ and $\sigma_{TT}$. The latter cancellations in the low-$Q$ region will be discussed in the following subsection.

\subsection{Low-$Q$ region and contribution of the $\bar S_1(0,Q^2)$ subtraction function} \seclab{ExpComp}

One major drawback of the data-driven dispersive evaluations is that  they require independent input for the inelastic spin structure functions or related polarizabilities, and the elastic Pauli form factor. 
Our notation in \Eqref{Delta1b} conveniently  illustrates how the zeroth moment of the inelastic spin structure function $g_1$ and the elastic Pauli form factor $F_2$ combine in the subtraction function:
\beq
\bar S_1(0,Q^2)=\frac{2\pi \al}{M}\left[F_{2}^2(Q^2)+4I_1(Q^2)\right]=\frac{8\pi \al}{M}\bar I_1(Q^2).\eqlab{S10def}
\eeq
At $Q^2=0$, this is zero, because the Pauli form factor, $F_2(0)=\kappa$, and the generalized GDH integral, $I_1(0)=-\kappa^2/4$, so the two terms cancel exactly. A NLO B$\chi$PT prediction of the slope amounts to: $[\ol I_1]'(0)=0.39(4)\,\mathrm{GeV}^{-2}$ \cite{Alarcon:2020icz}. It can be expressed through a combination of lowest-order spin [$\gamma_{E1M2}$] and generalized polarizabilities [$P^{\prime(M1,M1)1}(0)$ and 
$P^{\prime(L1,L1)1}(0)$], see \Eqref{S10Q2polexp}. In the HB$\chi$PT expansion, we showed that the leading $\mathcal{O}(\nicefrac{1}{m_\pi^2})$ terms cancel among these individual polarizabilities, given in \Eqref{individualPol}, turning the result subleading in $\mathcal{O}(\nicefrac{1}{m_\pi})$, see \Eqref{HBS1}. We can conclude that there is a strong cancellation between the elastic and inelastic contributions, which continues for higher $Q^2$. 

The contribution of $\bar S_1(0,Q^2)$ to the hfs is given by:
\beq
\eqlab{S1subtermhfs}
E_\mathrm{hfs}^{\langle \bar S_1(0,Q^2)\rangle }(nS)=\frac{E_\mathrm{F}}{n^3}
\frac{\al m}{\pi (1+\kappa) M}\int_0^\infty\frac{\dd Q}{Q}\frac{5+4v_l}{(v_l+1)^2}\left[4I_1(Q^2)+F_{2}^2(Q^2)\right].\qquad
\eeq
 Evaluations of this subtraction function contribution with empirical parametrizations for $g_1(x,Q^2)$ and $F_2(Q^2)$ tend towards larger values than the LO B$\chi$PT prediction. A partial calculation of the TPE effect at NLO in B$\chi$PT, considering only the one-loop box diagram with intermediate $\Delta(1232)$-excitation, will lower the theoretical prediction for the polarizability contribution from B$\chi$PT further, and in fact, turn it into a negative contribution \cite{Hagelstein:2018bdi,Hagelstein:2017cbl}. Any imprecision in the empirical parametrizations, and thus in the cancellation between the elastic and inelastic moments, is enhanced by the $1/Q$ prefactor in the infrared region of the integral in \Eqref{S1subtermhfs}. Therefore, the B$\chi$PT calculation, where the polarizability effect can be accessed directly through the non-Born part of the VVCS amplitudes and does not rely on input from separate measurements, has a clear advantage in this regard. 

To illustrate this further, we reproduce the estimate for $\Delta_1$ in the low-$Q$ region from Ref.~\cite{Carlson:2008ke} (see references therein for the details on the input). In this region, no experimental data from EG1 \cite{Prok:2008ev,Dharmawardane:2006zd} exist and the integral is completed by interpolating data between higher $Q^2$ and $Q^2=0$, making use of empirical values for the static polarizabilities.  For $Q^2 \in \left\{0,Q_\mathrm{max}^2\right\}$ with $Q_\mathrm{max}^2=0.0452\,\mathrm{GeV}^2$, the approximate formulas read  \cite{Carlson:2008ke}: 
\begin{subequations}
\eqlab{CarlsonDelta1estimate}
\bea
\delta_1(\text{H})&\sim&\left(\underbrace{-\frac{3}{4}\kappa^2 r_\mathrm{Pauli}^2}_{\rightarrow \,-2.19} +\underbrace{18M^2 c_{1B}}_{\rightarrow \,3.54} \right)Q_\mathrm{max}^2=1.35(90),\\
\delta_1(\mu\text{H})&\sim&\left[\underbrace{-\frac{1}{3}\kappa^2  r_\mathrm{Pauli}^2}_{\rightarrow\, -1.45} +\underbrace{8 M^2 c_{1}}_{\rightarrow\, 2.13} \underbrace{-\frac{M^2}{3 \alpha} \gamma_0}_{\rightarrow\, 0.18} \right]\int_0^{Q_\mathrm{max}^2} \dd Q^2 \beta_1(\tau_\mu)=0.86(69),\eqlab{approxmuHCarlson}
\eea
\end{subequations}
where
\beq
\beta_1(\tau_\mu)=-3 \tau_\mu+2 \tau_\mu^2 +2 (2-\tau_\mu) \sqrt{\tau_\mu  (\tau_\mu +1)}.
\eeq
Note that the formulas  for H and $\mu$H differ, because one sets $m_e=0$. 
 The first terms are related to the elastic Pauli form factor, where $r_\mathrm{Pauli}=-\nicefrac{6}{\kappa}\, \nicefrac{\dd}{\dd Q^2}\,F_2(Q^2)\vert_{Q^2=0}$ is the Pauli radius. The other terms are related to the $g_1$ contribution. Considering the more general \Eqref{approxmuHCarlson},
 they are defined through:
\begin{subequations}
\bea
I_1(Q^2)&=&M \int_{\nu_0}^\infty\frac{\dd \nu}{\nu^2}\,g_1(\nu,Q^2)=-\kappa^2/4+2M^2 c_1 Q^2+\mathcal{O}(Q^4),\\
\gamma_0&=&\frac{2\alpha}{M}\int_{\nu_0}^\infty \frac{\dd \nu}{\nu^4}\,g_1(\nu,0).
\eea
\end{subequations}
The strong cancellation between elastic and inelastic contributions, observed in \Eqref{CarlsonDelta1estimate}, can be a source of uncertainty.

\begin{figure}[tbh]
\centering
\includegraphics[width=0.5\columnwidth]{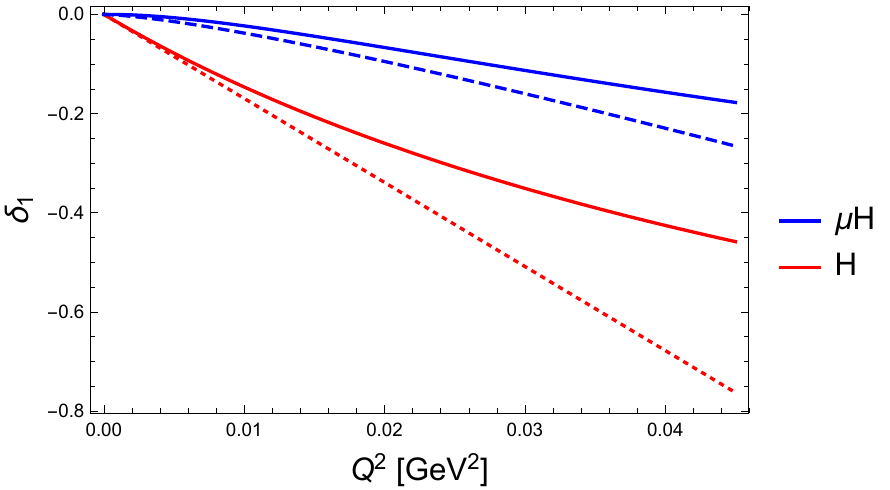}
\caption{The polarizability contribution $\delta_1$ in the low-$Q$ region for hydrogen (red) and muonic hydrogen (blue). The solid lines are the exact results according to \Eqref{Delta1b} with an upper cut on the $Q$ integration. The dotted and dashed lines are evaluated with the approximate formulas for hydrogen and muonic hydrogen, respectively, see \Eqref{CarlsonDelta1estimate}.}
\label{fig:LowQapprox}
\end{figure}

In addition, the quality of the low-$Q$ approximation is rather poor. We can test it at LO in B$\chi$PT. Recall that at this order in the B$\chi$PT power counting, there is no contribution to the elastic form factors. Therefore, only  the inelastic structure function $g_1$ enters. Our results are shown in \Figref{LowQapprox}. The approximate formulas in \Eqref{CarlsonDelta1estimate} give a $50\,\%$ ($67\,\%$) larger value for $\delta_1$ in the region of $Q^2<0.0452$ GeV$^2$ in the case of $\mu$H (H). Therefore, in the data-driven dispersive approach one has to properly account for the uncertainty introduced by the approximate formulas, as well as from cancellations between elastic and inelastic contributions.

\subsection{Scaling with lepton mass}\seclab{scalingSec}

It is customary to use the high-precision measurement of the $1S$ hfs in H \cite{Hellwig1970,Karshenboim:2000rg}:
\bea
E^{\,\text{exp.}}_{1S\text{-hfs}}(\mathrm{H})&=&1\,420.405\,751\,768(1)\,\text{MHz},\eqlab{expHHFS}
\eea
to refine the prediction of the TPE in the $\mu$H hfs \cite{Peset:2016wjq,Tomalak:2018uhr} or the prediction of the total $\mu$H hfs \cite{Antognini:2022xoo}. We will do the same in \secref{TPE}. The strategies in Refs.~\cite{Peset:2016wjq,Tomalak:2018uhr,Antognini:2022xoo} are slightly different, but all make statements about the scaling of various contributions to the hfs in a hydrogen-like atom when varying the lepton mass $m_\ell$. 

\begin{figure}[t]
\centering
\begin{minipage}{0.49\textwidth}
\centering
  \includegraphics[scale=0.52]{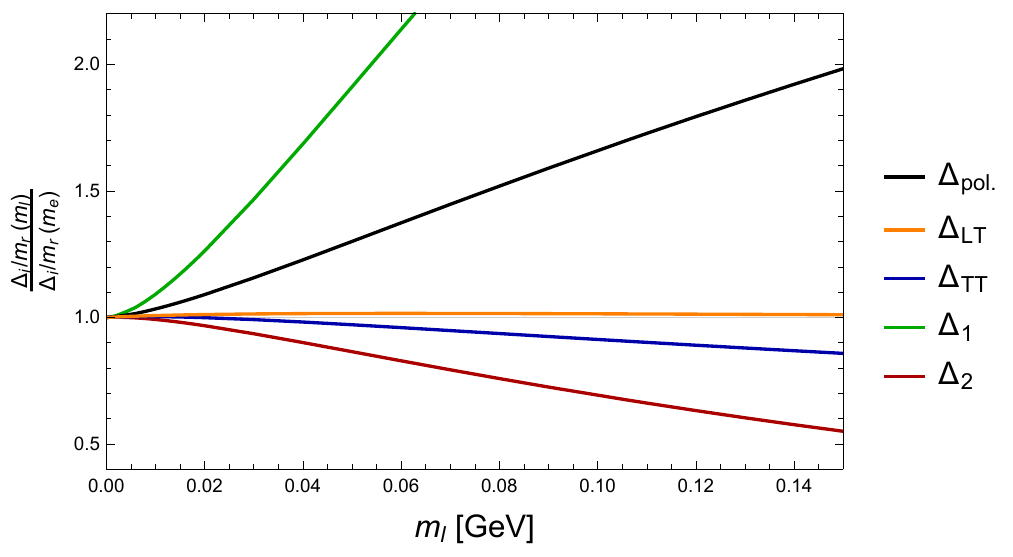}\vspace{0.25cm}
    \includegraphics[scale=0.52]{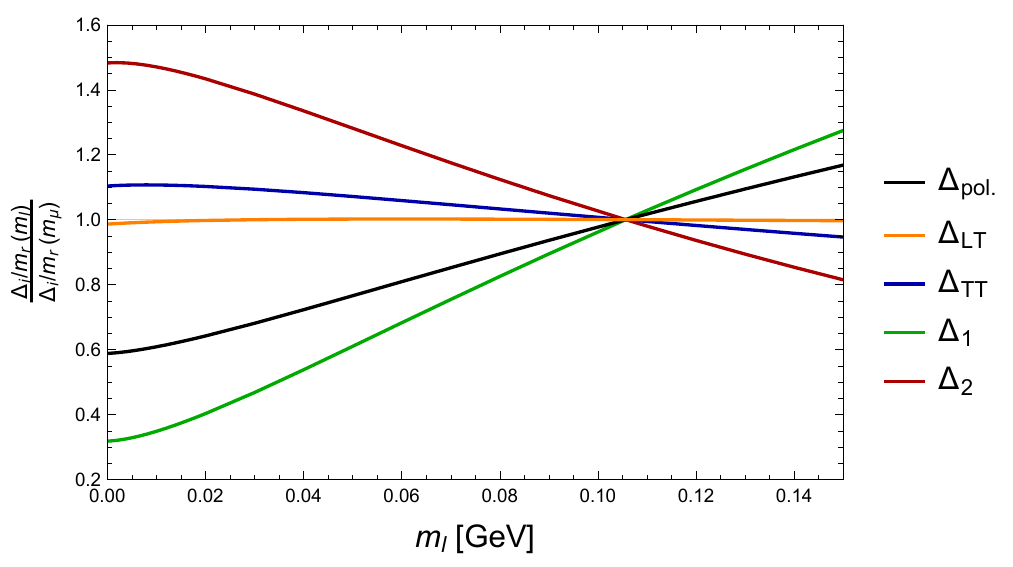}
\end{minipage}
\hfill
\begin{minipage}{0.49\textwidth}
\centering
       \includegraphics[scale=0.52]{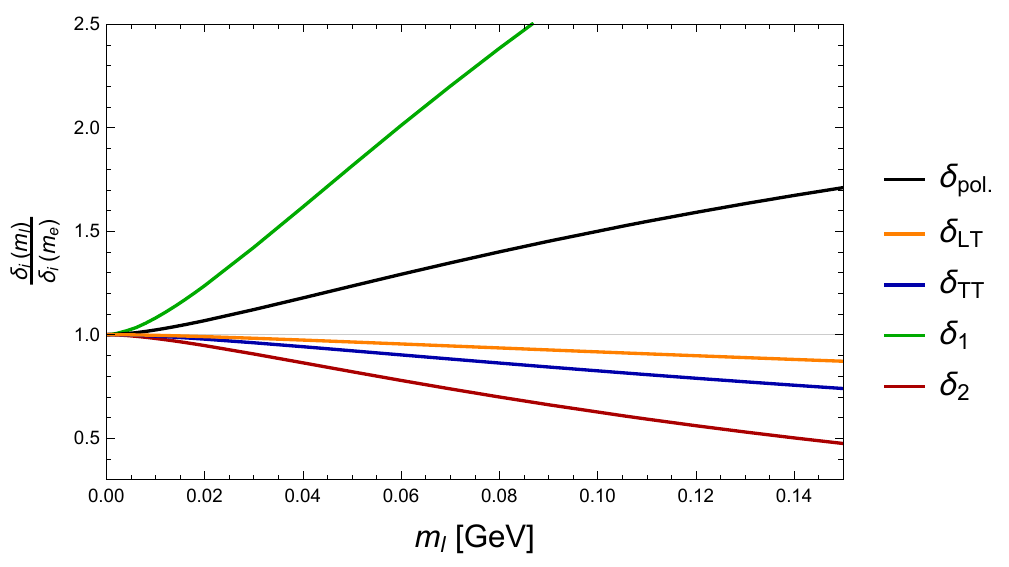}
       \includegraphics[scale=0.52]{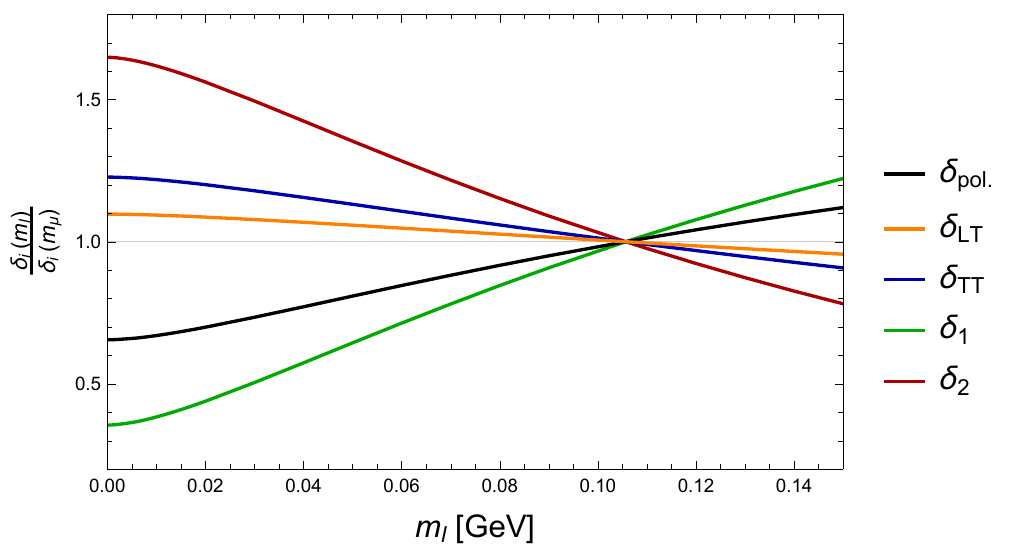}
    \end{minipage}
\caption{Scaling of $\delta_i$ and $\Delta_i/m_r$ (with $i=1,2,LT,TT$ and pol.), as a function of the lepton mass $m_\ell$. }
\label{Scaling}
\end{figure}

In Fig.~\ref{Scaling}, we study the scaling of the polarizability effect based on our LO B$\chi$PT prediction. In the left panel, we assume that the $\Delta_i$ (with $i=1,2,LT,TT$ and pol.) are scaling with the reduced mass $m_r$. In the right panel we assume that the  $\delta_i$ are independent of the lepton mass, thus, $\Delta_i$ would be scaling with $m_\ell$. The curves in the upper (lower) panel are normalized for H ($\mu$H), so they are fixed to $1$ at $m_\ell=m_e$ ($m_\ell=m_\mu$). If the polarizability effect would scale according to our assumptions, i.e., $\propto m_r$ or $\propto m_\ell$, all curves would be constantly $1$. We can see that the scaling works best for the contributions from $\sigma_{LT}$ and $\sigma_{TT}$, which are  large in their absolute values. Considering the total, in which the contributions from $\sigma_{LT}$ and $\sigma_{TT}$ cancel by about one order of magnitude, the scaling violation is enhanced by about one order of magnitude in relative terms. The same enhancement of the scaling violation can be observed for the numerically small contributions from $g_1$ and $g_2$. Comparing left and right panels, the B$\chi$PT predictions seems to support the assumption that $\Delta_{LT}$ and $\Delta_{TT}$ are scaling with $m_r$. For $\Delta_{LT}$, the scaling is nearly perfect. For $\Delta_{TT}$, we observe a violation of the scaling that is increasing with lepton mass. The approximation   $\Delta_{TT}(\mu\text{H}) \sim m_r(\mu\text{H}) / m_r(\text{H})\,\Delta_{TT}(\mu\text{H})$ holds at the level of  $10\%$. The approximation holds on a similar level after including an estimate for the NLO effect of the $\Delta(1232)$ resonance \cite{Hagelstein:2018bdi}.

\section{Extraction of the Zemach radius from spectroscopy} \seclab{ZemachRadius}

The TPE, entering the hfs, can be decomposed into Zemach radius, polarizability and recoil contributions, as described in \Eqref{FS_hfs}. On top of the $\mathcal{O}(\alpha^5)$ TPE, we consider the leading radiative corrections given by eVP, see \Figref{TPEwithVP} and discussion in Appendix \ref{eVPapp}.
Our prediction for the polarizability effect in the hfs, which is smaller than the conventional results from data-driven dispersive evaluations, also implies a smaller proton Zemach radius as previously determined from spectroscopy, cf.\ \Eqref{RZmuH}. In the following, we will extract the Zemach radius from the precisley measured $1S$ hfs in H, see \Eqref{expHHFS}, and the $2S$ hfs in $\mu$H \cite{Antognini:1900ns}:
\beq
E^{\,\text{exp.}}_\text{HFS}(2S, \mu\mathrm{H})=22.8089(51) \,\text{meV}.\eqlab{expmuHHFS}
\eeq
We use the theory predictions for the $1S$ hfs in H \cite{Antognini:2022xoo}:
\bea
E_{\text{hfs}}(1S, \text{H})&=&  \Big[1\,420\,453.106(10) \underbrace{-54.430(7)\,\left(\frac{R_{\mathrm{Z}}}{\text{fm}}\right)+E_\mathrm{F}\,\Big(0.99807(13)\,\Delta_\mathrm{recoil}+1.00002\,\Delta_\mathrm{pol.}\Big)}_{\text{TPE including radiative corrections}}\Big]\,\text{kHz} \eqlab{HFSHtheory}
\eea
and the $2S$ hfs in $\mu$H \cite{Antognini:2022xoo}:
\beq
\eqlab{2Spredic}
E_{\text{hfs}}(2S, \mu\text{H})= \Big[22.9584(8)\underbrace{-0.16319(2)\left(\frac{R_{\mathrm{Z}}}{\text{fm}}\right) +\frac{E_\mathrm{F}}{8}\,\Big(1.01580(4)\,\Delta_\mathrm{recoil}+1.00326\,\Delta_\mathrm{pol.}\Big)}_{\text{TPE including radiative corrections}}\Big]\,\text{meV}, 
\eeq
with the recently re-evaluated $\mathcal{O}(\al^5)$ recoil correction \cite{Antognini:2022xqf}:
\begin{subequations}
 \bea
\Delta_\mathrm{recoil}(\mathrm{H})&=&5.269\,{}^{+0.017}_{-0.004}\;\text{ppm},\\
\Delta_\mathrm{recoil}(\mu\mathrm{H})&=&837.6\,{}^{+2.8}_{-1.0}\;\text{ppm},
\eea
\end{subequations}
up to a factor $3$ more precise than the previous best determination \cite{Tomalak:2017lxo}
based on the electromagnetic form factors obtained from dispersion theory \cite{Lin:2021xrc}. An itemized list of contributions to the $2S$ hfs in $\mu$H is given in Table \ref{Table:Summaryhfs2S} of Appendix \ref{TheorySummary}.
From the LO B$\chi$PT prediction for the polarizability effect, including also the eVP in \Eqref{eVPnumbers}, we obtain:
\begin{subequations}
\eqlab{newRZs}
\bea
R_\mathrm{Z}(\mathrm{H})&=&1.010(9)\,\mathrm{fm},\eqlab{RZneweH}\\
R_\mathrm{Z}(\mu\mathrm{H})&=&1.040(33)\,\mathrm{fm}.\eqlab{RZnew}
\eea
\end{subequations}
This can be compared to other determinations of the proton Zemach radius collected in Table \ref{Table:ZemachRadius}.\footnote{Note that the chiral logarithm result for the Zemach radius \cite{Peset:2014jxa},
$R_\mathrm{Z}=1.35\,\mathrm{fm}$,
is substantially larger than all extractions from experiment.} The  radii we find are in agreement with the proton form factor analysis from Ref.~\cite{Borah:2020gte}, which uses the proton charge radius from the $\mu$H Lamb shift \cite{Antognini:1900ns} as a constraint for their fit.

\begin{figure}[t]
\centering
 \includegraphics[width=0.7\textwidth]{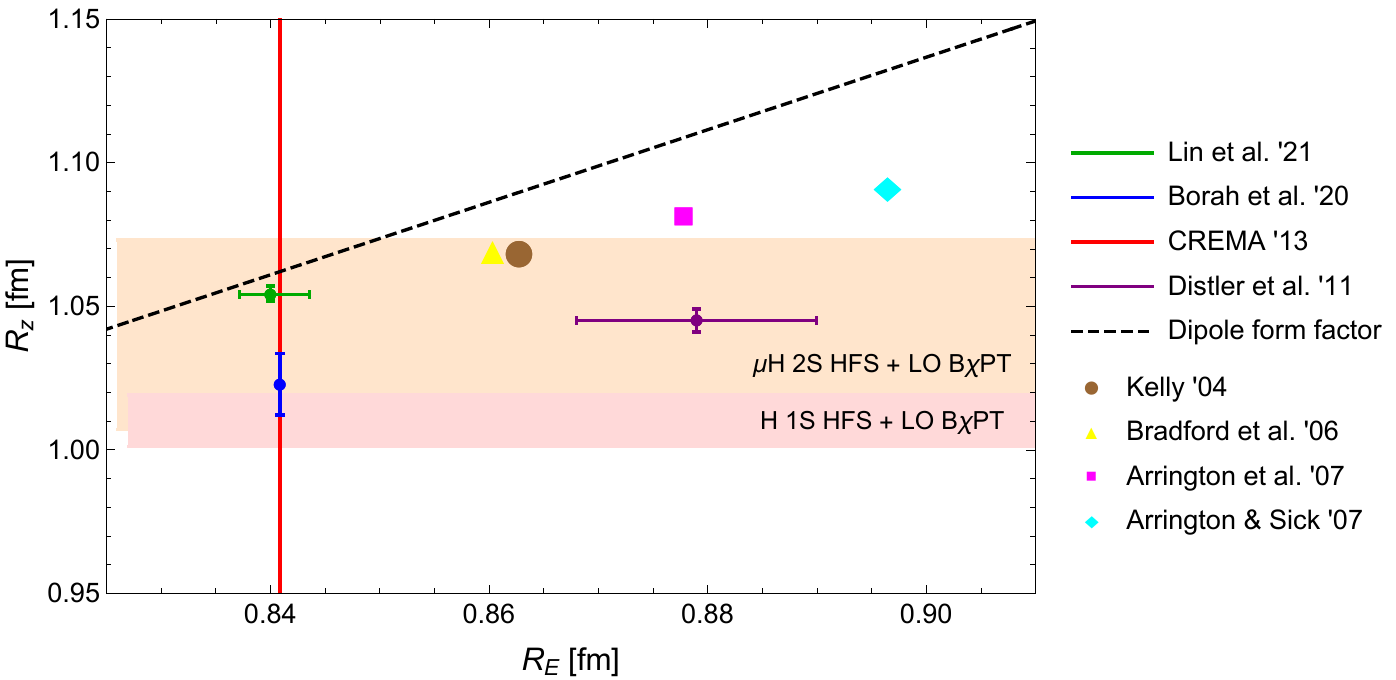}
\caption{Correlation between the Zemach and charge radius of the proton. Our extractions based on LO B$\chi$PT are compared to results from Lin et al.~\cite{Lin:2021xrc}, Borah et al.~\cite{Borah:2020gte}, CREMA \cite{Antognini:1900ns}, Distler et al.~\cite{Distler:2010zq}, Kelly \cite{Kelly:2004hm}, Bradford et al.~\cite{Bradford:2006yz}, Arrington et al.~\cite{Arrington:2007ux}, and Arrington \& Sick \cite{Arrington:2006hm}.
        \label{fig:RadiiCorrelation}}
\end{figure}

Figure~\ref{fig:RadiiCorrelation} shows how the Zemach and charge radius of the proton are correlated. It suggests that a ``smaller'' charge radius, as seen initially in the $\mu$H Lamb shift by the CREMA collaboration \cite{Antognini:1900ns} (red line), comes with a ``smaller'' Zemach radius.  The dashed black curve is calculated with a dipole form, $G(Q^2)\propto (1+Q^2/\Lambda^2)^{-2}$, for the electric and magnetic Sachs form factors, by varying $\Lambda$. The light red and orange bands show $R_Z$ as extracted by us, \Eqref{newRZs}, based on the LO B$\chi$PT prediction for the polarizability effect in the hfs.

\begin{table}[b]
\renewcommand{\arraystretch}{1.5}
\caption{Determinations of the proton Zemach radius $R_{\mathrm{Z}}$, in units of fm.}
\begin{footnotesize}
\label{Table:ZemachRadius}
\centering
\begin{tabular}{c|c|c|c|c|c}
\hline
\hline
\multicolumn{2}{c|}{$ep$ scattering}&\multicolumn{2}{c|}{$\mu$H $2S$ hfs}&\multicolumn{2}{c}{H $1S$ hfs}\\
\hline
Lin \textit{et al.} '21~\cite{Lin:2021xrc}
&
Borah \textit{et al.} '20~\cite{Borah:2020gte}
&
Antognini \textit{et al.} '13~\cite{Antognini:1900ns}
&
LO B$\chi$PT
&Volotka \textit{et al.} '04~
\cite{Volotka:2004zu}
&
LO B$\chi$PT 
\\
$1.054^{+0.003}_{-0.002}$&$1.0227(107)$&$1.082(37)$&$1.040(33)$&$1.045(16)$&$1.010(9)$\\
\hline
\hline
\end{tabular}
\end{footnotesize}
\end{table}
\renewcommand{\arraystretch}{1}

\section{Theory prediction for the ground-state hyperfine splitting in $\mu$H} \seclab{TPE}

The upcoming measurements of the $1S$ hfs in $\mu$H \cite{Amaro:2021goz,Pizzolotto:2021dai, Pizzolotto:2020fue,Sato:2014uza} crucially rely on a precise theory prediction. The limiting uncertainty is given by the TPE, which is conventionally split into Zemach radius, polarizability and recoil contributions
\cite{Antognini:2022xoo}:
\beq
\eqlab{1stheoryreview}
 E_{\text{hfs}}(1S, \mu\text{H})=  \Big[183.797(7)\underbrace{-1.30653(17)\left(\frac{R_{\mathrm{Z}}}{\text{fm}}\right)+ E_\mathrm{F}\,\Big(1.01656(4)\,\Delta_\mathrm{recoil}+1.00402\,\Delta_\mathrm{pol.}\Big)}_{\text{TPE including radiative corrections}}\Big]\,\text{meV},
 \eeq
 see Appendix \ref{TheorySummary} and Table \ref{Table:Summaryhfs1S} for 
an itemized list of the individual contributions. As explained in \secref{scalingSec}, it is customary to refine the theory prediction of $1S$ hfs in $\mu$H with the help of the high-precision measurement of the $1S$ hfs in H. We do so by combining
our B$\chi$PT prediction for the polarizability effect in the $\mu$H hfs, \Eqref{eVPnumbersmuH}, and the Zemach radius extracted from H spectroscopy, \Eqref{RZneweH}, based on the same prediction for the polarizability effect in the H hfs.
We arrive at:
\begin{subequations}
\bea
E_{\text{hfs}}(1S, \mu\text{H})&=&182.640(18)\,\mathrm{meV},\\
E^\text{TPE}_{\mathrm{hfs}}(1S, \mu\text{H})&=&-1.157(16)\,\mathrm{meV},
\eea
\end{subequations}
where $E^\text{TPE}_{\mathrm{hfs}}$ corresponds to the TPE including radiative corrections and recoil corrections from Ref.~\cite{Antognini:2022xqf}, as indicated by the curly brace in \Eqref{1stheoryreview}.

\begin{figure}[t]
\centering
\includegraphics[width=0.75\columnwidth]{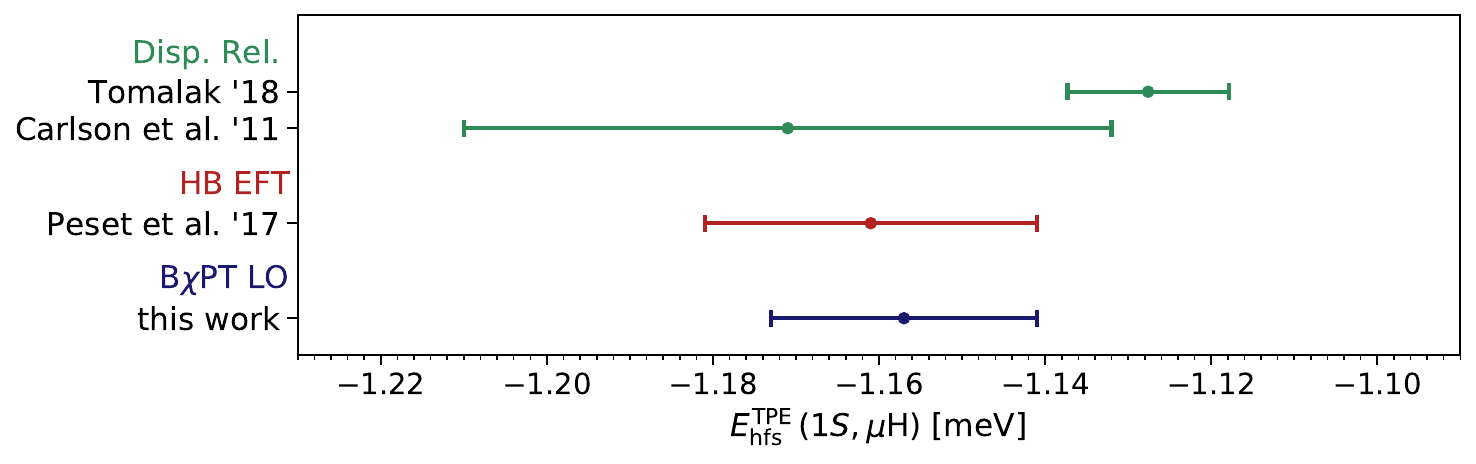}
\caption{Two-photon-exchange effect on the $1S$ hyperfine splitting in $\mu$H \cite{Carlson:2008ke,Peset:2016wjq,Tomalak:2018uhr}.}
\label{fig:TPEprediction}
\end{figure}

In Figs.~\ref{fig:TPEprediction} and \ref{fig:Totalprediction}, we compare our  predictions to results from data-driven dispersive evaluations \cite{Carlson:2008ke,Tomalak:2018uhr} and HB EFT \cite{Peset:2016wjq}.
While almost all available predictions for the total hfs in $\mu$H are in agreement after the H refinement procedure, further improvements of the theory are required in order to compete with the anticipated experimental accuracy.

\section{Conclusions and outlook}
\seclab{concl}

\begin{figure}[t]
\centering
\includegraphics[width=0.75\columnwidth]{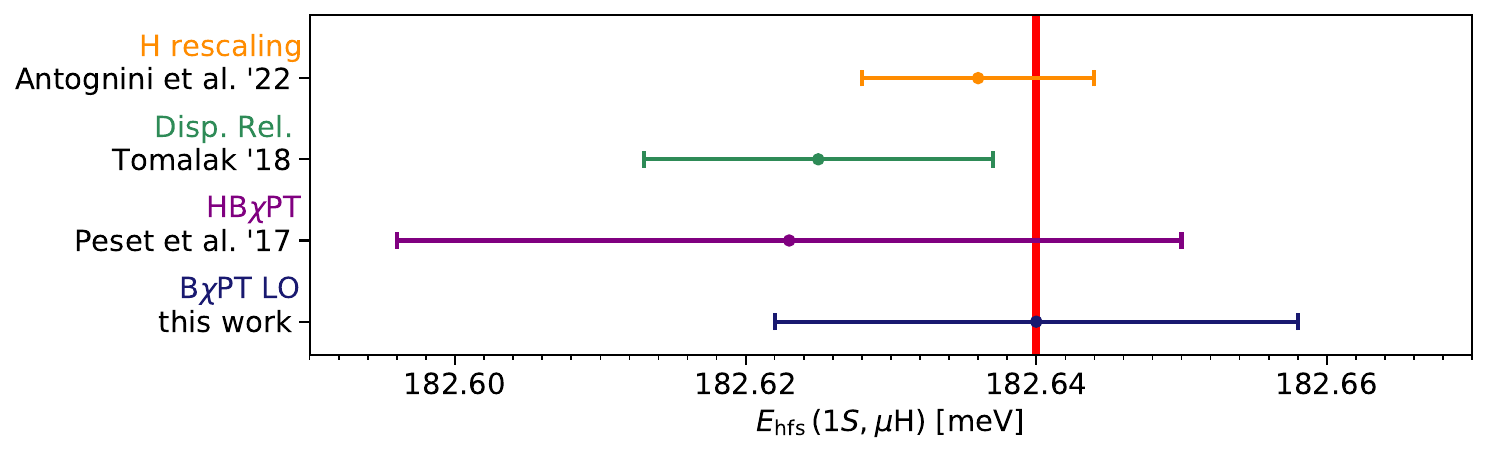}
\caption{Predictions for the $1S$ hyperfine splitting in $\mu$H \cite{Antognini:2022xoo,Peset:2016wjq,Tomalak:2018uhr}, compared to the \textit{projected uncertainty} of the planned CREMA measurement (red vertical line).}
\label{fig:Totalprediction}
\end{figure}

We have presented the LO B$\chi$PT prediction for the $\mathcal{O}(\al^5)$  polarizability effect on the hfs in H and $\mu$H, see \Eqref{pionresulthfs}. Contrary to the data-driven evaluations, the B$\chi$PT prediction is compatible with zero. This was expected from the HB$\chi$PT limit of the VVCS amplitudes, in particular $\bar S_1(0,Q^2)$, which partially display a cancellation of the leading order in the chiral expansion of small $m_\pi$, see discussion in
\secref{HBExpPion}. The small polarizability effect is then mainly a remnant of higher orders in the HB expansion. 

A new formalism where the polarizability effect is split into contributions from the longitudinal-transverse and helicity-difference cross sections,
$\sigma_{LT}$ and  $\sigma_{TT}$, instead of contributions from the spin structure functions, $g_1$ and  $g_2$,
has been introduced in \Eqref{POLAlternative}. It was shown that these contributions,
$\Delta_{LT}$ and  $\Delta_{TT}$, cancel by one order of magnitude when combined into $\Delta_\mathrm{pol.}$. 
Only $\Delta_{LT}$ and  $\Delta_{TT}$ are good observables in the B$\chi$PT framework, for which the contributions from beyond the scale at which this EFT is safely applicable, $Q_\mathrm{max}>m_\rho=775$ MeV, are within the expected uncertainty. In addition, only $\Delta_{LT}$ and  $\Delta_{TT}$ satisfy the conventionally assumed scaling with the reduced mass $m_r$ of the hydrogen-like system to $10\%$ relative accuracy, while the cancellations in $\Delta_\mathrm{pol.}$ enhance any violation in the scaling by one order of magnitude.

As shown in Fig.~\ref{fig:CompEH}, 
our model-independent LO B$\chi$PT prediction is substantially smaller than the data-driven dispersive evaluations. 
 An estimate for the effect of the $\Delta(1232)$-resonance \cite{Hagelstein:2018bdi}, obtained from large-$N_c$ relations for the nucleon-to-delta transition form factors, shows that the discrepancy is likely to increase at the NLO. The smaller polarizability effect, in turn, leads to a smaller Zemach radius as extracted from  the experimental $1S$ hfs in H and the $2S$ hfs in $\mu$H, cf.\ \Eqref{newRZs}. Therefore, resolving the present discrepancy for the polarizability effect is crucial for the analysis of the forthcoming measurements of the $1S$ hfs in $\mu$H and the extraction of the Zemach radius. 
 
The data-driven approach relies on empirical information on the inelastic spin structure functions, or the measured cross sections to be precise, as well as the elastic form factors and polarizabilities at $Q^2=0$. Due to the large cancellations between $\sigma_{LT}$ and $\sigma_{TT}$, as well as $g_1$ and $F_2$, precise parametrizations of the former are needed, and the uncertainty of the TPE evaluation has to be estimated with great care, taking into account all correlations. Furthermore, due to a lack of data at low-$Q$, one uses an interpolation from $Q^2=0$ to the onset of data \cite{Carlson:2008ke}. As we showed in \secref{ExpComp} based on LO B$\chi$PT, the quality of these approximations is rather poor and is yet another source of uncertainty.
New data from the Jefferson Lab ``Spin Physics Program'' \cite{Chen:2008ng,CLAS:2017ozc,CLAS:2021apd,JeffersonLabE97-110:2019fsc,E97-110:2021mxm}, including also the substantially extended dataset for $g_2$ \cite{JeffersonLabHallAg2p:2022qap}, will allow for a re-evaluation of the polarizability effect on the hfs in H and $\mu$H.

 An accurate theoretical prediction of the $1S$ hfs in $\mu$H is crucial for the future measurement campaigns, since it allows to reduce the search range for the resonance in experiment. Thus, one might find the resonance faster and acquire more statistics during the allocated beam time, see discussion in Ref.~\cite{Antognini:2022xoo}. The present discrepancy between predictions for the polarizability effect can be mended if the high-precision measurement of the $1S$ hfs in H is implemented as a constraint. Applying this procedure, good agreement is found between all theory predictions for the total $1S$ hfs in $\mu$H hfs, see \Figref{Totalprediction}. Eventually, after a successful measurement of the $1S$ hfs in $\mu$H, one can
combine it with the $1S$ hfs in H to disentangle the Zemach radius and polarizability effects, leveraging radiative corrections as explained in Ref.~\cite{Antognini:2022xoo}. The empirical polarizability effect, obtained in this way, can reach a precision of $\sim 40$ ppm \cite{Antognini:2022xoo}. That is sufficient to discriminate between the presently inconsistent theoretical predictions.

\section*{Acknowledgements}
   We thank   A.~Antognini, C.~Carlson, J.\ M.\ Alarc\'on, and A.~Pineda for useful discussions, S.~Simula for providing a {\sc FORTRAN} code with his latest parametrization of the spin-dependent proton structure functions, and L.~Tiator for explanations on the MAID isobar model. 
This work is supported by the Deutsche Forschungsgemeinschaft (DFG) 
through the Emmy Noether Programme grant 449369623, the Research Unit FOR5327 grant 458854507, and partially by the Swiss National Science Foundation (SNSF) through the Ambizione Grant PZ00P2\_193383.

\appendix

\section{Two-photon-exchange master formula and dispersive approach}
\label{AppendixTPE}

The proton-structure effects at $\mathcal{O}(\al^5)$ are described by TPE in forward kinematics, i.e., by the diagram in \Figref{TPE} where the momentum transfer between the initial and final particles is vanishing. The (forward) TPE can be related to the amplitudes of (forward) VVCS off the proton, which in turn can be expressed in terms of proton structure functions via dispersion relations. A detailed review of the VVCS theory can be found in Ref.~\cite[Section 5]{Hagelstein:2015egb}. Even though the TPE formalism is well-known, see for instance Ref.~\cite{Carlson:2008ke,Carlson:2011af}, we will present here its derivation for the hfs.\footnote{An extensive discussion of the TPE formalism, considering in addition the Lamb shift, can also be found in Ref.~\cite[Chapter 5]{Hagelstein:2017cbl}.}

As implied above, it is customary to split the TPE into leptonic and hadronic tensors ($L^{\mu \nu}$ and $T^{\mu \nu}$):
\bea
\mathcal{M}&=&\frac{1}{2}\!\int \!\frac{\dd^4q}{i(2\pi)^4}\frac{1}{q^4}\left[\bar{u}(\ell)L_{\mu \nu}(\ell,q)u(\ell)\right]\left[\bar{N}(p)T^{\mu \nu}(p,q)N(p)\right]\!,
\eqlab{TPEM}
\eea
where $u(\ell)$ and $N(p)$ are the lepton and proton Dirac spinors, with $\ell$, $p$ and $q$ being the lepton, proton and photon four-momenta (see \Figref{TPE}), respectively. Here, a factor of $\nicefrac{1}{2}$ has been introduced to avoid double counting when contracting the crossing-invariant tensors.

Only the spin-dependent part of the forward VVCS will contribute to the hfs. For the proton, it reads:\footnote{We define $\ga_{\mu\nu}=\half\left[\ga_\mu,\ga_\nu\right]$ and $\ga_{\mu\nu\al}=\half(\ga_\mu\ga_\nu\ga_\al - \ga_\al\ga_\nu\ga_\mu)$.}
\bea
T^{\mu \nu}_A(q,p)&=& -\frac{1}{M}\gamma^{\mu \nu \al} q_\al \,S_1(\nu, Q^2) 
+\frac{Q^2}{M^2}  \gamma^{\mu\nu} S_2(\nu, Q^2),
\eqlab{VVCS_TA}
\eea
where $S_1$ and $S_2$ are two independent scalar functions of the photon lab-frame energy $\nu$ and the photon virtuality $Q^2 =-q^2= \bq^2 -\nu^2$. Equivalently, one can write  \Eqref{VVCS_TA} with the help of 
the spin four-vector $\mathtt{s}^\al$ (satisfying $\mathtt{s}^2=-1$ and $\mathtt{s}\cdot p=0$):\footnote{The  symmetric spin-independent part of forward VVCS reads:
\beq
%\eqlab{TS}
T^{\mu \nu}_S(q,p) = -g^{\mu\nu}\,
T_1(\nu, Q^2)  +\frac{p^{\mu} p^{\nu} }{M^2} \, T_2(\nu, Q^2),\nonumber
\eeq
where terms vanishing upon contraction with the lepton tensor are not shown.}
\bea
T^{\mu \nu}_A (q,p) &=& 
\frac{i}{M}\,\epsilon^{\mu\nu\alpha\beta}\,q_{\alpha}
\mathtt{s}_{\beta}\, S_1(\nu, Q^2)+ \frac{i}{M^3}\,\epsilon^{\mu\nu\alpha\beta}\,q_{\alpha}
(p\cdot q\ \mathtt{s}_{\beta}-\mathtt{s}\cdot q\ p_{\beta})\, S_2 (\nu, Q^2).
\eea
Since $T_A^{\mu \nu}$ is antisymmetric in its indices, it is sufficient to replace the lepton tensor with the antisymmetric part of the tree-level QED amplitude of forward VVCS [in the structureless limit with the Dirac and Pauli form factors $F_1\rightarrow 1$ and $F_2\rightarrow 0$, cf. Eqs.~\eref{VVCS_TA} and \eref{S12Born}]:\footnote{The symmetric part of the tree-level QED amplitude of forward VVCS, contributing to the Lamb shift, reads:\beq
L_S^{\mu \nu}=\frac{4\pi \al }{m}\frac{1}{(\ell\cdot q)^2-\quarter Q^4}\left[g^{\mu\nu}(\ell\cdot q)^2-(q^\mu\ell^\nu+q^\nu\ell^\mu)\,\ell\cdot q-Q^2 l^\mu l^\nu\right].%\eqlab{LeptonTensorLS}
\nonumber
\eeq}
\beq
L_A^{\mu \nu}=\frac{-2\pi \al \,Q^2}{(\ell\cdot q)^2-\quarter Q^4}\,\gamma^{\mu\nu \al }q_\al.\eqlab{LeptonTensor}
\eeq

The lepton and proton momenta are of typical atomic scales, thus, much smaller than the other scales we are considering. Therefore, in the center-of-mass frame, we assume both of them to be at rest, $p=\nicefrac{M}{m} \,\ell=(M,\boldsymbol{0})$:\footnote{%Angular averaging gives: $\int \dd \Omega_q\,\, \hat{q}_i\hat{q}_j=\nicefrac{1}{3}\,\delta_{ij}$, for a unit 3-vector $\hat q$.
Averaging over the angles of $\bq$ gives $\left(\bq\cdot\boldsymbol{s}\right)\left(\bq\cdot\boldsymbol{S}\right)\to \nicefrac{1}{3}\, \bq^2 \left(\boldsymbol{s}\cdot\boldsymbol{S}\right)$ and so on for other combinations. }
\begin{align}
&\hspace{-0.1cm}\left[\bar{u}(\ell)\gamma_{\mu \nu\al} q^\al\,u(\ell)\right]\left[\bar{N}(p)\gamma^{\mu \nu}\,N(p)\right]= 8 \nu \,\boldsymbol{s} \cdot \boldsymbol{S},\qquad\\
&\hspace{-0.1cm}\left[\bar{u}(\ell)\gamma_{\mu \nu\al} q^\al u(\ell)\right]\left[\bar{N}(p)\gamma^{\mu \nu \beta} q_\beta N(p)\right]= \frac{8}{3}(\nu^2-2Q^2)\,\boldsymbol{s} \cdot \boldsymbol{S},\nn
\end{align}
with the lepton and proton spin operators $\boldsymbol{s}$ and $\boldsymbol{S}$, and $\ell \cdot q=m \nu$.

The forward TPE generates a $\delta(\vec{r}\,)$-function potential: $V(r)=\mathcal{M}\,\delta(\vec{r}\,)$.
Treated in perturbation theory, such kind of potential generates an energy shift of the $nS$ levels: $E_{nS}= \phi_n^2 \,\mathcal {M}$,
where $\phi_n^2=1/(\pi a^3 n^3)$ is the hydrogen wave function at the origin. When acting on the wave functions, the product of spin operators can be replaced by the atom's total angular momentum $f$ \cite{BetheSalpeter}: 
\beq
\boldsymbol{s} \cdot \boldsymbol{S}\xrightarrow{{nS\text{ level }}}\half\left[f(f+1)-\mbox{$\frac{3}{2}$}\right]. \eqlab{sShfs}
\eeq
Recalling that the $nS$ hfs is defined as the splitting between $S$ levels with $f=1$ and $f=0$, we finally arrive at the following master formula for the TPE contribution to the hfs in terms of proton VVCS amplitudes:
\bea
E^\mathrm{TPE}_{\mathrm{hfs}}(nS)&=&\frac{E_\mathrm{F}}{n^3}\frac{4m}{1+\kappa}\frac{1}{i}\int_{-\infty}^\infty \!\frac{\dd\nu}{2\pi} \int \!\!\frac{\dd \bq}{(2\pi)^3}\,\frac{1}{Q^4-4m^2\nu^2}\left\{\frac{\left(2Q^2-\nu^2\right)}{Q^2}S_1(\nu,Q^2)+\frac{3\nu}{M}S_2(\nu,Q^2)\right\}.\eqlab{VVCS_hfs} 
\eea
The Lamb shift analogue reads
\beq
E^\mathrm{TPE}_{nS}= 8\pi \al m \,\phi_n^2\,
\frac{1}{i}\int_{-\infty}^\infty \!\frac{\dd\nu}{2\pi} \int \!\!\frac{\dd \bq}{(2\pi)^3}\frac{\left(Q^2-2\nu^2\right)T_1(\nu,Q^2)-(Q^2+\nu^2)\,T_2(\nu,Q^2)}{Q^4(Q^4-4m^2\nu^2)},\eqlab{LSMaster}
\eeq
where $T_1$ and $T_2$ are the spin-independent forward VVCS amplitudes.
The VVCS amplitudes can be split into Born and non-Born parts:
\begin{subequations}
\bea
S_1(\nu,Q^2)&=&S_1^{\mathrm{Born}}(\nu, Q^2)+\bar S_1(\nu,Q^2),\\
S_2(\nu,Q^2)&=&S_2^{\mathrm{Born}}(\nu, Q^2)+\bar S_2(\nu,Q^2),
\eea
\end{subequations}
where Born corresponds to the simplest tree-level diagrams with a proton in the intermediate state, and non-Born corresponds to everything else. The finite-size recoil and Zemach radius effects are described by the well-known Born part of the VVCS amplitudes \cite{Drechsel:2002ar}:
\begin{subequations}
\eqlab{S12Born}
\bea
S_1^{\mathrm{Born}}(\nu, Q^2) &=& \frac{2\pi \alpha}{M}
\bigg[\frac{4M^2 Q^2\,G_{M}(Q^2)F_{1}(Q^2)}{Q^4-4M^2\nu^2}-F_{2}^2(Q^2)\bigg], \eqlab{S1Born}\\
S_2^{\mathrm{Born}}(\nu, Q^2)&=&
-\, \frac{8 \pi \alpha M^2 \nu}{Q^4-4M^2\nu^2}\,G_{M}(Q^2) F_{2}(Q^2),
\eqlab{S2Born}
\eea
\end{subequations}
where $F_{1}(Q^2)$ and $F_{2}(Q^2)$ are the Dirac and Pauli form factors of the proton, and $G_{M}(Q^2)=F_{1}(Q^2)+F_{2}(Q^2)$ is the magnetic Sachs form factor. The Zemach radius and its effect on the hfs are defined in Eqs.~\eref{ZemachTerm} and \eref{RZdef}. Exact formulas for the TPE recoil contribution can be found for instance in Ref.~\cite{Carlson:2011af}. The polarizability effect is described by the non-Born part of the VVCS amplitudes.

Using the general principles of analyticity, unitarity, crossing symmetry and gauge invariance, one can conveniently express the spin-dependent VVCS amplitudes through the proton spin structure functions $g_1(x,Q^2)$ and $g_2(x,Q^2)$ by means of the optical theorem:
\begin{subequations}
\eqlab{LTTT}
\bea
\im S_1(\nu,Q^2) &=& \frac{4\pi^2 \alpha}{\nu} \, g_1(x,Q^2) = 
\frac{M  \nu^2 }{\nu^2+Q^2}\left[\frac{Q}{\nu}\sigma_{LT}  + \sigma_{TT}\right](\nu,Q^2), \eqlab{ImS1}\\
\im S_2(\nu,Q^2) & =&  \frac{4\pi^2 \alpha M}{\nu^2} \, g_2(x, Q^2)  
= \frac{M^2\nu}{\nu^2+Q^2}\left[\frac{\nu}{Q}\sigma_{LT}  - \sigma_{TT}\right](\nu,Q^2), \qquad \quad\eqlab{ImS2}
\eea
\end{subequations}
and dispersion relations:
\begin{subequations}
\eqlab{S12nB}
\bea
\bar S_1(\nu,Q^2)&=&\frac{2\pi \al}{M} \Bigg\{ \left[F_{2}^2(Q^2)+4I_1(Q^2)\right]+\frac{32 M^4\nu^2}{Q^6}  \int_{0}^{x_0} 
\!\dd x\,
\frac{x^2 g_1 (x, Q^2)}{1 - x^2 (\nu/\nu_{\mathrm{el}})^2  - i 0^+}\Bigg\},\eqlab{S1subtrDR}\\
\nu \bar S_2(\nu,Q^2)&=&\frac{64\pi \al M^4\nu^2}{Q^6}  \int_{0}^{x_0} 
\!\dd x\,
\frac{x^2 g_2 (x, Q^2)}{1 - x^2 (\nu/\nu_{\mathrm{el}})^2  - i 0^+},\eqlab{S2subtrDR}
\eea
\end{subequations}
with $\nu_{\mathrm{el}}=\nicefrac{Q^2}{2M}$ and $I_1(Q^2)$ defined in \Eqref{I1def}.
As shown in \Eqref{LTTT}, the spin structure functions correspond to certain combinations of photoabsorption cross sections. Here, $\sigma_{LT}$ is the longitudinal-transverse photoabsorption cross section describing a spin-flip of the proton, and $\sigma_{TT}=\nicefrac12\, (\sigma_{1/2}-\sigma_{3/2})$ is the helicity-difference cross section for transversely polarized photons, where the subscripts on $\sigma_{1/2}$ and $\sigma_{3/2}$ denote the total helicity of the $\gamma^\ast N$ state. 

To solve \Eqref{VVCS_hfs} one uses a Wick rotation to imaginary energies, and hyperspherical coordinates. Employing the dispersive representation from \Eqref{S12nB}, after the angular integrations one obtains the  polarizability effect as presented in \Eqref{POL}, where one conventionally splits into contributions from $g_1$ and $g_2$. As we explain in our paper, in view of the uncertainty estimate, it is favorable to consider instead a splitting into contributions from $\sigma_{LT}$ and $\sigma_{TT}$, derived by us in \Eqref{POLAlternative}. 

It is worth to discuss some subtleties entering the definition of the polarizability effect through the non-Born VVCS amplitudes in the dispersive approach. Naively, one would expect the Born and non-Born amplitudes to be expressed entirely through elastic form factors and inelastic structure functions, respectively. Instead, one finds:
\begin{subequations}
\bea
S_1^{\mathrm{elastic}}(\nu, Q^2)-S_1^{\mathrm{Born}}(\nu, Q^2) &=&\bar S_1(\nu, Q^2)-S_1^{\mathrm{inelastic}}(\nu, Q^2)= \frac{2\pi \al}{M}F_2^2(Q^2)\eqlab{S1corr},\\
\left[\nu S_2\right]^{\mathrm{elastic}}(\nu, Q^2)- \nu S_2^{\mathrm{Born}}(\nu, Q^2) &=&\nu \bar S_2(\nu, Q^2)-\left[\nu S_2\right]^{\mathrm{inelastic}}(\nu, Q^2)=- 2\pi \al\,F_2(Q^2)G_M(Q^2),\eqlab{S2corr}
\eea
\end{subequations}
where $S_1^{\mathrm{elastic}}$ and $\left[\nu S_2\right]^{\mathrm{elastic}}$ are pure nucleon-pole terms. In
\Eqref{S1subtrDR}, the necessary conversion term to obtain the non-Born amplitude, given on the right-hand side of \Eqref{S1corr}, is easily seen. Here,
even though $S_1$ satisfies an unsubtracted dispersion relation, we wrote a once-subtracted dispersion relation, with the subtraction term $\bar S_1(0,Q^2)$ defined in \Eqref{S10def}. This is useful to emphasize the interplay of the elastic Pauli form factor $F_2$ and the inelastic spin structure function $g_1$, as explained in 
Sections \ref{sec:Calculation} and \ref{sec:ExpComp}. For \Eqref{S2subtrDR}, the applied conversion procedure is less obvious. Starting from a dispersion relation for $\nu S_2$:\footnote{The amplitude $S_2$ does have a pole in the subsequent limit of $Q^2\rightarrow 0$ and $\nu\rightarrow 0$.}
\beq
\nu S_2 ( \nu, Q^2) =2\pi \al \,\frac{2}{\tau}  \int_{0}^1 
\!\dd x\, 
\frac{g_2 (x, Q^2)}{1 - x^2 (\nu/\nu_{\mathrm{el}})^2  - i 0^+}, 
\eeq
we separate the inelastic part (i.e., limit the integration to $x \in \{0,x_0\}$) and add the conversion term from the right-hand side of \Eqref{S2corr}. The latter  then cancels the inelastic part of the zeroth moment of the $g_2$ structure function, due to the
Burkhardt-Cottingham (BC) sum rule \cite{Burkhardt:1970ti}: 
\beq
0 = \int_{0}^1\dd x\,  g_{2}(x,\,Q^2)=\int_{0}^{x_0}\dd x\,  g_{2}(x,\,Q^2) -\frac{\tau}{2}F_2(Q^2)G_M(Q^2),
\eqlab{BCsumrule}
\eeq
leading to \Eqref{S2subtrDR}. Thus, splitting into Born and non-Born amplitudes, the BC sum rule constraint is automatically satisfied. In other words, we showed that:
\beq
\nu S_2^{\mathrm{Born}}(\nu, Q^2)\big\vert_{\nu \rightarrow 0} =\nu \bar S_2(\nu, Q^2)\big\vert_{\nu \rightarrow 0}=0.
\eeq

\section{Electron vacuum polarization correction}\label{eVPapp}

\begin{figure}[t]
\includegraphics[width=0.15\textwidth]{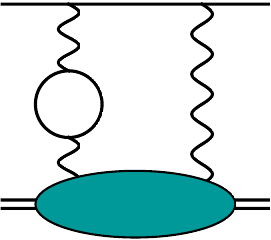}
\caption{Two-photon exchange with vacuum-polarization insertion at $O(\al^6)$. \figlab{TPEwithVP}}
\end{figure}

In this appendix, we consider the one-loop eVP correction to the TPE, shown in \Figref{TPEwithVP}. This amounts to multiplying the integrand in \Eqref{VVCS_hfs} with $\left[1-\ol \Pi^{(1)}(Q^2)\right]^{-2}$,
where the VP is given by:
\begin{align}
    \ol \Pi^{(1)}(Q^2)= \Pi^{(1)}(Q^2)-\Pi^{(1)}(0) & = \frac{\alpha}{3\pi}  \left[2 \left(1-\frac{1}{2\tau_e }\right) \left(\sqrt{1+\frac{1}{\tau_e }}
    \arccoth\sqrt{1+\frac{1}{\tau_e }}
    -1\right)+\frac{1}{3}\right],
\end{align}
with $\tau_e=Q^2/4m_e^2$ and $m_e$ the electron mass. The resulting corrections to the polarizability effect are given in \Eqref{eVPnumbers}.

\renewcommand{\arraystretch}{1.3}
\begin{table}[t]
%\begin{sidewaystable}[t]
\caption{$1S$ hfs in $\mu$H. All values in meV}
\label{Table:Summaryhfs1S}
\begin{tabular}{c|c|r|c}
\hline
\hline
&Contribution&Our Choice&Reference\\
\hline
h1&Fermi energy, $(Z\al)^4$&$182.44333$&\\
h2&Breit corr., $(Z\al)^6$&$0.01457$&\\
h4&$\mu$ anomalous magnetic moment corr., $\al(Z\al)^4$&$0.21271$&\\
\hline
h5&eVP in 2\textsuperscript{nd}-order PT, $\al^2(Z\al)^4 m_r$&$0.73449$& \cite[Table 1 b)]{Karshenboim2009} 
\\
h7&Two-loop corr.\ to Fermi energy, $\al^2(Z\al)^4 m_r$&$0.00556$& \cite[Table 2 c) and d)]{Karshenboim2009} \\
\hline
h8&One-loop eVP in $1\gamma$ int., $\al(Z\al)^4 m_r$ &$0.37465$&\cite[Table 1 a)]{Karshenboim2009} 
\\
h9&Two-loop eVP in $1\gamma$ int., $\al^2(Z\al)^4 m_r$&$0.00292$& \cite[Table 2 a) and b)]{Karshenboim2009} \\
h10&Further two-loop eVP corr. in 2\textsuperscript{nd} and 3\textsuperscript{rd}-order PT &$0.00387$&\cite[Table 2 e), f) and g)]{Karshenboim2009}\\
\hline
 h11&$\mu$VP&$0.00729$&$\sim E_\mathrm{F}\,\alpha (Z\alpha) \,3/4$\\
\hline
h13&Vertex, $\al(Z\al)^5$&$-0.02484$&$\sim E_\mathrm{F}\,\alpha (Z\alpha) \left[\ln 2-13/4\right]$ \\
h14&Higher-order corr.\  $\al(Z\al)^6$&$-0.00128$& \cite[Eq.~7.1]{Brodsky:1966vn}\\
\hline
h18&hVP, $\al^6$&$0.00356$&\cite{Faustov:1997rc}\\
h19&Weak interaction contribution&$0.00221$&\cite[Eq.~374]{Eides:2000xc}\\
\hline
h28&Recoil corr.\ with $p$ AMM, $\al^6$&$0.01752$&\cite[Eq.~22]{Carlson:2008ke} and \cite{Bodwin:1987mj}\\
\hline
\hline
\end{tabular}
\end{table}
\renewcommand{\arraystretch}{1}

\renewcommand{\arraystretch}{1.3}
\begin{table}[t]
\caption{$2S$ hfs in $\mu$H. All values in meV}
\label{Table:Summaryhfs2S}
\begin{tabular}{c|c|r|c}
\hline
\hline
&Contribution&Our Choice&Reference\\
\hline
h1&Fermi energy, $(Z\al)^4$&$22.80542$&\\
h2&Breit corr., $(Z\al)^6$&$0.00258$&\\
h4&$\mu$ anomalous magnetic moment corr., $\al(Z\al)^4$&$0.02659$&\\
\hline
h5&eVP in 2\textsuperscript{nd}-order PT, $\al^2(Z\al)^4 m_r$&$0.07447$& \cite[Table 1 b)]{Karshenboim2009} 
\\
h7&Two-loop corr.\ to Fermi energy, $\al^2(Z\al)^4 m_r$&$0.00056$& \cite[Table 2 c) and d)]{Karshenboim2009} \\
\hline
h8&One-loop eVP in $1\gamma$ int., $\al(Z\al)^4 m_r$ &$0.04828$&\cite[Table 1 a)]{Karshenboim2009} 
\\
h9&Two-loop eVP in $1\gamma$ int., $\al^2(Z\al)^4 m_r$&$0.00037$& \cite[Table 2 a) and b)]{Karshenboim2009} \\
h10&Further two-loop eVP corr. in 2\textsuperscript{nd} \& 3\textsuperscript{rd}-order PT &$0.00037$&\cite[Table 2 e), f) and g)]{Karshenboim2009}\\
\hline
 h11&$\mu$VP&$0.00091$&$\sim E_\mathrm{F}\,\alpha (Z\alpha) \,3/4$\\
\hline
h13&Vertex, $\al(Z\al)^5$&$-0.00311$&$\sim E_\mathrm{F}\,\alpha (Z\alpha) \left[\ln 2-13/4\right]$ \\
h14&Higher-order corr.\  $\al(Z\al)^6$&$-0.00013$& \cite[Eq.~7.1]{Brodsky:1966vn}\\
\hline
h18&hVP, $\al^6$&$0.0006(1)$&\cite{Borie:2012zz}\\
h19&Weak interaction contribution&$0.00028$&\cite[Eq.~374]{Eides:2000xc}\\
\hline
h21&Higher-order finite-size corr.\ to Fermi energy&$-0.0022\, r_p^2+0.0009$&\cite[Eq.~107]{Indelicato:2012pfa}\\
&&$\approx -0.00065$&\\
\hline
h28&Recoil corr.\ with $p$ AMM, $\al^6$&$0.00185$&\cite[Eqs.~1.3 and 2.13]{Peset:2016wjq}\\
\hline
\hline
\end{tabular}
\end{table}

\section{Theory compilation for hyperfine splitting in $\mu$H}\label{TheorySummary}

In this appendix, we present the details of our theory compilations for the $1S$ and $2S$ hfs in $\mu$H, shown in Eqs.~\eref{2Spredic} and \eref{1stheoryreview}. All individual contributions, except the TPE, are listed in Tables \ref{Table:Summaryhfs1S} and \ref{Table:Summaryhfs2S}. The notation is the same as in Ref.~\cite{Antognini:2012ofa}.   The difference with Ref.~\cite{Peset:2021iul} is in the inclusion of \#\#h10, h14, h18, h19 and h21, as well as some additional radiative corrections to the TPE evaluated in Ref.~\cite{Antognini:2022xoo}.  Compared to Refs.~\cite{Antognini:2012ofa,Borie:2012zz}, we suggest to use Eq.~(7.1) from  Ref.~\cite{Brodsky:1966vn}:\footnote{Note that some terms in Eqs. (C5) and (C9) of that reference appear to be missing the Fermi energy factor.} 
\bea
E^{[\al(Z\al)^6]}_{nS\text{-hfs}}&=&\frac{\al(Z\al)^2}{\pi}\,\frac{E_\mathrm{F}}{n^3} \left\{-\frac{8}{3}\ln^2 \frac{2n}{Z\al}+\left[\frac{37}{36}+\frac{8}{15}+7(n-1)\right]\ln \frac{n}{2Z\al}\right.\nn\\
&&\left.+\left[\frac{22}{3}\ln 2-\frac{2\pi^2}{9}+18-\frac{457}{2700}-\left(4+\frac{2993}{8640}\right)(n-1)\right]\right\},   \nn
\eea
for the higher-order corrections of $\mathcal{O}(\al(Z\al)^6)$ [\#h14].
This includes higher-order muon vacuum polarization corrections. Previously included were only the 
 logarithmically enhanced terms \cite{Borie:2012zz}.
The effect on the TPE from eVP corrections to the wave function is given in Ref.~\cite[Eq.~(B3)]{Karshenboim:2021jsc}.
For the radius-independent term, we are keeping the error estimate from Ref.~\cite{Peset:2016wjq}, which does  take into account missing higher-order recoil corrections.

\section{Expansions in terms of polarizabilities}

In the following, we will present two further low-energy expansions of the polarizability effect in the hfs. Due to the high-energy asymptotics of the TPE contribution to the hfs, these formulas will merely serve illustrative purposes, while their approximation of the full result is rather poor.
Up to and including second moments of the structure functions, \Eqref{POL} can be written as \cite{Hagelstein:2017cbl}:

\begin{subequations}
\eqlab{polD1}
\bea
\delta_1&=&2\int_0^\infty\frac{\dd Q}{Q}\frac{1}{(v_l+1)^2}\bigg\{4(5+4v_l)\,\bar{I}_1(Q^2)-\frac{11+9v_l}{(v_l+1)}\left[\frac{ M^2Q^2}{2\al}\,\gamma_0(Q^2)+\frac{32Z^2M^6}{Q^6}\int_0^{x_0}\dd x\, x^4\, g_2(x,Q^2)\right]\!\bigg\},\qquad\quad\eqlab{polD1b}\\
\delta_2
&=&-24\int_0^\infty\frac{\dd Q}{Q}\frac{1}{(v_l+1)^2}\left\{\frac{ M^2Q^2}{2\al}\,\left[\delta_{LT}(Q^2)-\gamma_0(Q^2)\right]-\frac{32Z^2M^6}{Q^6}\int_0^{x_0}\dd x\, x^4\, g_2(x,Q^2)\right\},\eqlab{polD2b}
\eea
\end{subequations}
where $\ga_0(Q^2)$ and $\delta_{LT}(Q^2)$ are the forward spin and longitudinal-transverse polarizabilities of the proton, 
\begin{subequations}
\bea
\gamma_0(Q^2)&=&\frac{16 \al M^2}{Q^6}\int_0^{x_0}\dd x\, x^2 \Big[g_1-x^2\tau^{-1}g_2\Big](x,Q^2)=\frac{1}{2\pi^2}\int_{\nu_0}^\infty\frac{\dd \nu}{\nu^3}\,\sigma_{TT}(\nu,Q^2),\\
\delta_{LT}(Q^2)&=&\frac{16Z^2 \al M^2}{Q^6}\int_0^{x_0}\dd x\, x^2 \big[g_1+g_2\big](x,Q^2)=\frac{1}{2\pi^2}\int_{\nu_0}^\infty  \frac{\dd \nu}{Q\nu^2}\,\sigma_{LT}(\nu,Q^2).
\eea
\end{subequations}
The first term in this expansion, corresponds to the $\bar S_1(0,Q^2)$ subtraction term already discussed in \secref{ExpComp}. 

\begin{figure}[bht]
\centering
\includegraphics[width=0.5\columnwidth]{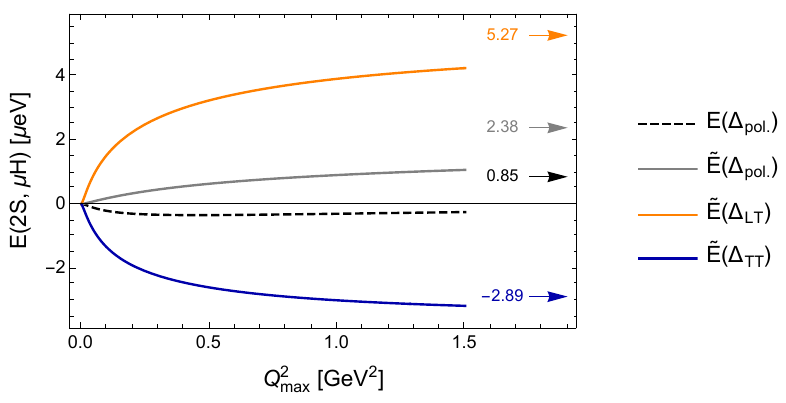}
\caption{Polarizability effect on the $2S$ hyperfine splitting in $\mu$H: Comparison of exact result \eref{POL} [black dashed line] and approximate formula \eref{Approxhfs} [gray solid line].  }
\label{HBfig}
\end{figure}

Analogously to Ref.\ \cite[Eq.~(12)]{Alarcon:2013cba}, we try to find an approximation for the hfs master formula assuming that the photon energy in the atomic system is small compared to all other scales. Thus, we expand the numerator of \Eqref{VVCS_hfs} around $\nu=0$. The resulting approximate formula for the polarizability contribution to the hfs we call $\tilde E$:
\beq
\eqlab{Approxhfs}
\tilde E(nS)=\frac{E_\mathrm{F}}{n^3}\frac{4\al}{\pi mM}\frac{1}{1+\kappa}\int_{0}^\infty \!\dd Q\, Q\, (v_l-1)\,\bar I_1(Q^2).
\eeq
For B$\chi$PT, it gives:
\begin{subequations}
\bea
\tilde E(1S, \mathrm{H})&=&3.0\,\text{peV},\\
\tilde E(1S, \mu\mathrm{H})&=&19.0 \,\upmu\text{eV}.
\eea
\end{subequations}
In Fig.~\ref{HBfig}, we show \Eqref{Approxhfs} as a running integral with cut-off $Q_\mathrm{max}$ (gray line), this time for the $2S$ hfs in $\mu$H. In addition, we show the contributions of the longitudinal-transverse (orange line) and helicity-difference (blue line) cross sections to \Eqref{Approxhfs}, and the exact result from \Eqref{POL}. One can easily see that the quality of the approximation is indeed rather poor. 

\renewcommand{\arraystretch}{1}

\bibliography{lowQ}

\end{document}